\documentclass[journal,twocolumn]{IEEEtran}
\usepackage[cmex10]{amsmath}
\usepackage{amssymb}
\usepackage{cite}
\usepackage{graphicx}
\usepackage{array,color}
\usepackage{multirow}
\usepackage{amsmath}
\usepackage{stfloats}
\usepackage{graphicx}
\usepackage{subfigure}
\usepackage{tabularx}
\usepackage{epsfig,epsf,color,balance,cite}
\usepackage{setspace}
\usepackage{bm}
\usepackage{textcomp}
\usepackage{algorithmic}
\usepackage{algorithm}

\usepackage{marginnote}

\usepackage{comment}
\usepackage{subfigure}

\usepackage{caption}
\usepackage{graphicx}
\usepackage{setspace}
\usepackage{diagbox}

\usepackage{multirow}
\usepackage{graphicx}
\makeatletter

\newcommand{\Rmnum}[1]{\expandafter\@slowromancap\romannumeral #1@}
\makeatother
\usepackage{booktabs}
\usepackage{amsmath}

\makeatother
\pdfoutput=1
\begin{document}
	
	\title{Near-Field Multiuser Beam-Training for Extremely Large-Scale  MIMO Systems}
	\author{
		Wang Liu, Cunhua Pan, $\textit{Senior Member, IEEE}$, Hong Ren, $\textit{Member, IEEE}$, Jiangzhou Wang, $\textit{Fellow, IEEE}$, Robert Schober, $\textit{Fellow, IEEE}$, and Lajos Hanzo, $\textit{Life Fellow, IEEE}$\thanks{
			\emph{Corresponding author: Cunhua Pan and Hong Ren.}
			
			Wang Liu, Cunhua Pan and Hong Ren are with National Mobile Communications Research Laboratory, Southeast University, Nanjing 210096, China. (e-mail: {w\_liu, cpan, hren}@seu.edu.cn).
			
			Jiangzhou Wang is with the School of Engineering, University of Kent, CT2 7NZ Canterbury, U.K. (e-mail: j.z.wang@kent.ac.uk).
			
			Robert Schober is with the Institute for Digital Communications, Friedrich-Alexander-Universität (FAU) Erlangen-Nürnberg, 91058 Erlangen, Germany (e-mail: robert.schober@fau.de).
			
			Lajos Hanzo is with the School of Electronics and Computer Science, University of Southampton, SO17 1BJ Southampton, U.K. (e-mail: lh@ecs.soton.ac.uk).
			
	} }

\maketitle
\begin{abstract}
Extremely large-scale multiple-input multiple-output (XL-MIMO) systems are capable of improving spectral efficiency by employing far more antennas than conventional massive MIMO at the base station (BS). However, beam training in multiuser XL-MIMO systems is challenging. Firstly, new near-field channel models and near-field XL-MIMO transmit beamforming (TBF) codebooks have to be adopted due to the dramatic increase in the number of antennas, which results in an excessive pilot overhead for beam training. Secondly, when the user density is high, the wireless propagation environments of the adjacent users are similar and hence the pilot signals received by the BS from different users appear to be interrelated, which is potentially beneficial but difficult to exploit. Thirdly, different users might share the same beam-direction, which causes excessive inter-user interference. To tackle these issues, we conceive a three-phase graph neural network (GNN)-based beam training scheme for multiuser XL-MIMO systems. \textit{In the first phase}, only far-field wide beams have to be tested for each user and the GNN is utilized to map the beamforming gain information of the far-field wide beams to the optimal near-field beam for each user. In addition, the proposed GNN-based scheme can exploit the position-correlation between adjacent users for further improvement of the accuracy of beam training. \textit{In the second phase}, a beam allocation scheme based on the probability vectors produced at the outputs of GNNs is proposed to address the above beam-direction conflicts between users. \textit{In the third phase}, the hybrid TBF is designed for further reducing the inter-user interference. Our simulation results show that the proposed scheme improves the beam training performance of the benchmarks based on traditional neural networks. Hence it is more suitable for multiuser XL-MIMO systems. Moreover, the performance of the proposed beam training scheme approaches that of an exhaustive search, despite requiring only about 7\% of the pilot overhead.
\end{abstract}

\vspace{-0.4cm}
\section{Introduction}
In the fifth-generation (5G) wireless network, massive multiple-input multiple-output (MIMO) techniques are harnessed for improving the spectral efficiency by employing massive antenna arrays at the base station (BS)\cite{5g_mimo_1,5g_mimo_2,5g_mimo_3}. However, the improvement in spectral efficiency provided by massive MIMO cannot satisfy the requirements of the sixth-generation (6G) wireless network in terms of spectral efficiency, which is tenfold higher compared to the 5G network\cite{6g_1,6G_3}. To meet the dramatically increased spectral efficiency requirements, extremely large-scale MIMO (XL-MIMO) employing many more antennas than massive MIMO, has received extensive attention \cite{xlmimo}. More explicitly, the XL-MIMO system provides both an improved signal-to-noise ratio (SNR) and spectral efficiency by creating high-gain directional beams at the BS. 
In order to enable accurate beamforming, codebook-based transmit beamforming (TBF) training schemes are widely adopted \cite{cui_1}, in which the BS tests the TBF weight-vectors in a predefined codebook and selects the one having the maximum beamforming gain. However, beam training for XL-MIMO systems faces many challenges, especially in multiuser scenarios.

\textit{Firstly}, the beam training of numerous antennas in the XL-MIMO systems would impose excessive pilot overheads. Specifically, XL-MIMO leads to a substantial shift of the boundary between the far-field and the near-field regions, i.e., the Rayleigh distance. As a result, the specifics of near-field communication have to be considered, because a large fraction of the users are likely to be located in the near-field region\cite{han_yu_1}. For near-field communications, a near-field codebook specifically designed for spherical wave-front must be adopted, which contains many more codewords than the conventional far-field codebooks \cite{cui_1}. Hence again, beam training in XL-MIMO systems imposes both excessive pilot overhead and high computational complexity, which makes our problem much more challenging than that of massive MIMO systems. \textit{This represents a qualitative rather than just quantitative difference.}
	
\textit{Secondly}, when the user density is high, the pilot signals received by the BS from different users appear to be interrelated, which is potentially beneficial but hard to exploit. In 6G systems having a high user-density, especially in massive communication scenarios proposed as usage scenario in \cite{ITU}, the users tend to be close to each other, hence they would have similar wireless propagation environments as well as similar system parameters such as noise intensity, path loss, scatterers and reflections. Additionally, users having similar locations would also have similar line-of-sight (LoS) and non-line-of-sight (NLoS) paths. Accordingly, the uplink (UL) pilot signals received by BS contain information about the wireless propagation environment, system parameters and paths. In a single-user scenario, the user transmits its UL pilot signals to the BS and then the BS determines the optimal TBF codeword purely based on the user's own pilot signals. By contrast, in the multi-user scenario, the users transmit orthogonal UL pilot signals to the BS. Not only do the user's own pilot signals help the BS determine the optimal TBF codeword, but the pilot signals of the surrounding users may also be beneficially exploited as ``copies" to provide diversity gain. Thus they assist the BS to determine the optimal codeword more accurately. However, how to extract and effectively exploit the information hidden in the pilot signals of the surrounding users is an open challenge, because the influence of the environment and system parameters on the pilot signals is an extremely complex issue.
	
\textit{Thirdly}, in the multi-user XL-MIMO systems, beam conflict becomes more obvious and it can cause severe inter-user interference due to the high user-density. When the distribution of users is more dense, the optimal beams for users in similar locations are likely to be the same. However, if the BS assigns the same beam for different users, severe inter-user interference will occur, which is also known as beam conflict. Therefore, in multi-user XL-MIMO systems, the problem of beam conflicts has to be addressed.

\vspace{-0.2cm}
\subsection{State-of-the-art}
\vspace{-0.2cm}
In conventional far-field communications, various beam training schemes were proposed for improving the accuracy \cite{limin}, for reducing the overhead of beam training \cite{hussain,h_codebok_1,h_codebok_2,qi_2} and for mitigating beam conflicts \cite{sun}. The authors of \cite{limin} proposed a two-stage search scheme that improves the accuracy of beam training under a fixed energy budget, but the pilot overhead is not reduced. Furthermore, an optimal interactive beam alignment scheme was proposed in \cite{hussain}, which reduces the power consumption and satisfies specific power constraints. Hierarchical codebook-based schemes were proposed in \cite{h_codebok_1,h_codebok_2} to reduce the pilot overhead required for beam training. In these schemes, wide-beam codewords having wider-angle coverage are used to roughly determine the range of optimal beam codewords at the outset. However, beam training schemes based on hierarchical codebooks require a large amount of feedback. Furthermore, the authors of \cite{qi_2} conceived a sophisticated solution where all users could rely on the same hierarchical beam training codebook, which reduced the amount of feedback required. In contrast to the above work, the beam training scheme proposed by the authors of \cite{sun} considered multiuser scenarios and reduced the inter-user interference, but the pilot overhead is still high. 

Moreover, beam training schemes employing popular deep learning methods were proposed in \cite{qi_1,make1,make4,heng_2} to incorporate powerful learning capabilities. The authors of\cite{qi_1} proposed a fully connected neural network (FCNN) based beam training scheme, where only a fraction of the codewords has to be tested and the corresponding received signals are used as input for the DNN. Furthermore, the authors of \cite{make1} proposed to employ efficient convolutional neural networks (CNN) for determining the optimal narrow-angle beam, while testing only the wide-angle beam codewords. To further reduce the pilot overhead, the authors of \cite{make4} proposed to utilize the pilot signals received at the sub-6G BS to narrow down the range of the optimal beam. In \cite{heng_2}, a beam training scheme utilising the user's 3D coordinates was proposed, where a FCNN was trained to map the coordinates into optimal beam codewords, but the acquisition of the user's coordinates is difficult in practice. However, all these contributions only take into account far-field channels and codebooks, but they cannot be readily generalised to near-field scenarios.

For the near-field domain, several schemes have been proposed to reduce the pilot overheads\cite{luyu,my_letter,jiangguoli}. Recently, the authors of \cite{luyu} proposed a beam training scheme based on a near-field hierarchical codebook, in which a wide-beam codeword suitable for the near-field channel was designed, but the proposed codebooks only considered single-user scenarios. In \cite{my_letter}, a CNN-based near-field beam training scheme was proposed, in which only far-field wide-beam codewords have to be tested, and then the result is entered into the trained CNN and mapped to the optimal near-field beams. Although this reduces the pilot overhead, the scheme proposed in \cite{my_letter} still has shortcomings that prevent it from employment in multiuser scenarios. The most important impediment is that the neural network structure used in \cite{my_letter} cannot exploit the pilot signals gleaned from the surrounding users. Furthermore, determining the optimal beam for each user directly based on the output of the neural network would result in severe beam conflicts. Similarly, the authors of \cite{jiangguoli} proposed a scheme for determining the optimal near-field beam based on the test information of partial near-field codewords, which adopted a CNN and a FCNN. However, the scheme still has the same drawbacks as the arrangement of \cite{my_letter} and it cannot be readily employed in multi-user scenarios. To the best of our knowledge, beam training for multi-user XL-MIMO systems has not yet been studied.

\vspace{-0.2cm}

\subsection{Main Contributions}
\vspace{-0.1cm}


To fill this gap, we propose a three-phase graph neural network (GNN) based beam training scheme for multi-user XL-MIMO systems, where the pilot overhead is reduced and the correlation of users' pilot signals is exploited. Furthermore, a matching beam allocation scheme is conceived for reducing beam conflicts. We also adopt practical hybrid precoding architectures, where a digital transmit precoder (TPC) is designed to further mitigate inter-user interference. Although the FCNN and CNN proposed in \cite{my_letter} and \cite{jiangguoli} have been shown to be effective in reducing the pilot overhead, a further contribution of this paper is the conception of a GNN-based beam training scheme for multiuser scenarios in order to better exploit the correlation of the users' pilot signals. Firstly, GNNs have a similar structure as FCNN and also possess powerful nonlinear mapping capabilities \cite{overview_gnn}. The authors of \cite{jiang_tao_1} adopted GNN to design the active precoding matrix of the BS and the reflection phase matrix at the reconfigurable intelligent surface (RIS), acting as a passive beamformer, which achieved an excellent performance. Secondly, we will demonstrate that the GNN can effectively exploit the pilot signals of the surrounding users and thus improve the efficiency of beam training. The GNN proposed in \cite{jiang_tao_1} was only used for modelling the interference between users, while ignoring the potential diversity gain gleaned from surrounding users. We go further and propose a GNN architecture for exploiting the pilot signals of surrounding users. Moreover, this meritorious GNN is not required to be retrained when the number of users changes, so it can be applied to any number of users. Additionally, existing neural network-based beam training schemes do not mitigate beam conflicts. Hence, we further propose a beam allocation scheme based on the probability vectors produced by the GNN to reduce beam conflicts and make full use of the probability vector for improving the accuracy of beam training.  Our main contributions can be summarized as follows:

\begin{enumerate}
	\item We propose a GNN-based beam training scheme relying on three phases for reducing the pilot overhead required for beam training in multi-user XL-MIMO systems, where the BS only has to test a comparatively small fraction of far-field wide-beam codewords for each user. 
	\item We show that in a multi-user scenario, the proposed GNN can substantially improve the estimation accuracy of the optimal near-field beam of a user by exploiting the pilot signals of the surrounding users. 
	\item We propose a beam allocation scheme based on probability vectors, which effectively mitigate beam conflicts. In this scheme, the probability vectors are further exploited, and thus the accuracy of beam training is improved.
	\item We provide extensive simulation results to characterize the performance of our proposed scheme. Our simulation results reveal that the proposed scheme outperforms the existing beam training schemes based on common neural network models. In addition, our proposed scheme achieves a similar performance to that of the exhaustive search, despite requiring only 7\% of the associated pilot overhead.
\end{enumerate}

The remainder of this paper is organized as follows. In Section \ref{systemmodel1}, the system model and the problem formulation are described. In Section \ref{DL_model}, the pilot transmission scheme and the architecture of GNN are provided. Additionally, the proposed three-phase GNN-based beam training scheme is presented. In Section \ref{simulation}, our simulation results are provided followed by our conclusions in Section \ref{conclusion}.

In this paper, we adopt the following notations. Vectors and matrices are represented by bold lower case and bold upper case letters, respectively, e.g., $\mathbf{a} $ and $\mathbf{A} $, while $a $ and $\mathcal{A}  $ denote a scalar and a set, respectively; $\left ( \cdot  \right )^{\ast }$, $\left ( \cdot  \right )^{\mathrm{T} }$ and $\left ( \cdot  \right )^{\mathrm{H} }$ represent conjugate, transpose, and conjugate transpose, respectively; $\left|\cdot  \right|$ denotes the absolute value; $\left [\mathbf{a}  \right ]_{i}$ denotes the $i$-th element of $\mathbf{a}$, and $\left [ \mathbf{A}  \right ] _{i,j}$ denotes the $(i,j)$-th element of $\mathbf{A}$; $\mathcal{C} \mathcal{N}(\mu ,\sigma^{2} )$  represents the Gaussian distribution with mean $\mu$ and variance $\sigma^{2}$. $\left \|  \cdot \right \| _{\mathrm{F} } $ denotes the Frobenius norm.

\vspace{-0.4cm}

\section{System Model}\label{systemmodel1}

\vspace{-0.2cm}
\subsection{Signal Model}

\begin{figure}[t]
	\centering
	\includegraphics[width=3.5in]{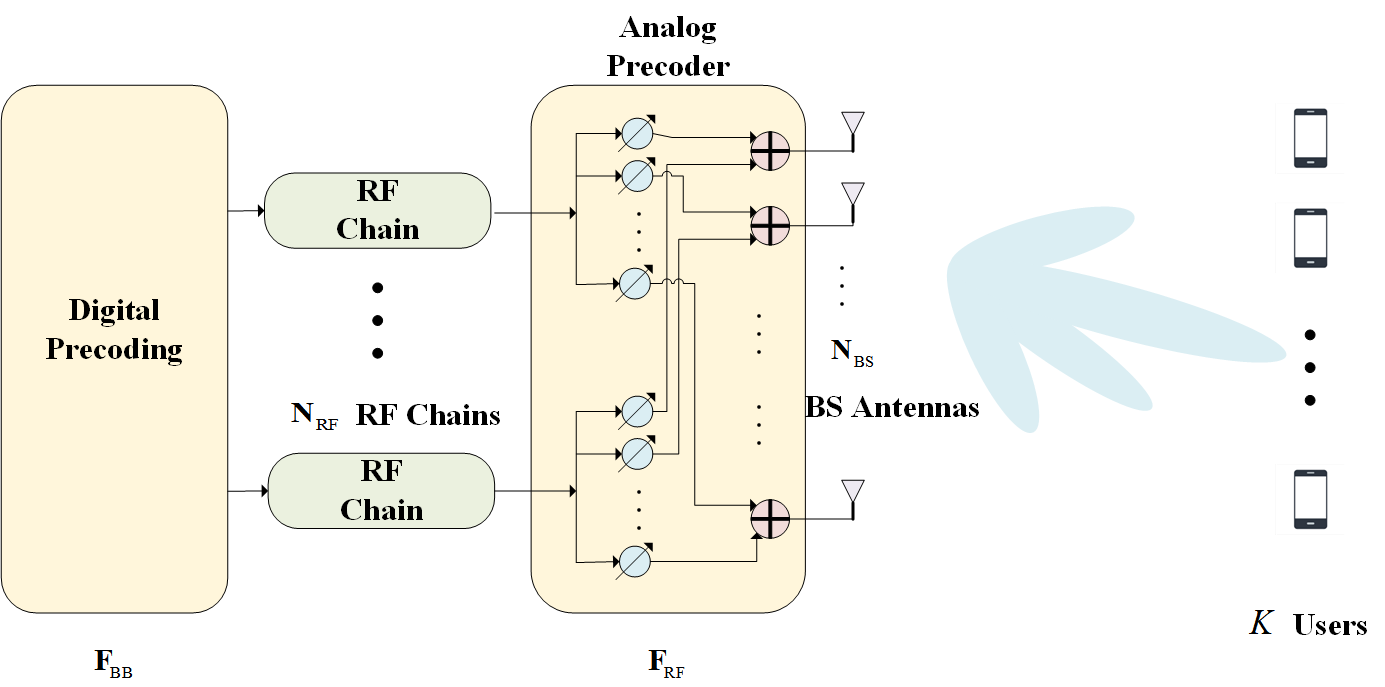}
	\caption{Hybrid precoding architecture for multiuser XL-MIMO system. }
	\label{hp}\vspace{-0.8cm}
\end{figure}

We consider a time division duplex (TDD) based multi-user mmWave communication system, as shown in Fig.\ref{hp}, where $K$ single-antenna users are served by a BS employing a uniform linear array (ULA) with $N_{\textrm{BS}}$ antennas\footnote{	Note that our proposed scheme can be readily extended to the systems having uniform planer arrays (UPA). When a UPA is considered, the proposed scheme only requires replacing the 2-dimensional near-field codebook with the 3-dimensional near-field codebook, and the network used to determine the angle index becomes the networks used to determine the azimuth angle index and the elevation angle index, respectively.}. 
The BS is equipped with $N_{\textrm{RF}}$ radio frequency (RF) chains and a hybrid TPC architecture to enable simultaneous communication with the $K$ users, where $K\leq N_{\textrm{RF}} \ll N_{\textrm{BS}}$ is satisfied. To save power consumption, redundant RF chains will be turned off and $N_{\textrm{RF}}=K$ is satisfied \cite{sun,hybrid_precoding_AA_1}. 

Let us denote the downlink (DL) channel from the BS to the $k$-th user by $\mathbf{h}_{k}^{\mathrm{dl}}\in \mathbb{C}^{1\times N_{\textrm{BS}}}$. During DL transmission, the signal received by the $k$-th user can be represented as\cite{hybrid_precoding_AA_1,hybrid_precoding_AA_2,sun}
\begin{equation}\label{dl_r}
	\setlength\abovedisplayskip{3pt}
	\setlength\belowdisplayskip{3pt}
	\begin{aligned}
		{{r}}_k^{{\rm{dl}}} &= {\bf{h}}_k^{{\rm{dl}}}{{\bf{F}}_{{\rm{RF}}}}{{\bf{F}}_{{\rm{BB}}}}{\bf{s}} + n_k^{{\rm{dl}}}\\
		&= {\bf{h}}_k^{{\rm{dl}}}{{\bf{F}}_{{\rm{RF}}}}\sum\limits_{n = 1}^K {{\bf{f}}_n^{{\rm{BB}}}{s_n} + n_k^{{\rm{dl}}}},
	\end{aligned}
\end{equation}
where $\mathbf{s}=\left [ s_{1},s_{2}, \cdots ,s_{K}\right ]^{\textrm{T}}$ denotes the signal vector transmitted by the BS, which is subject to a total downlink power constraint, i.e., $\mathbb{E}\left [ \mathbf{s} \mathbf{s}^{\textrm{H}}\right ]=\frac{P_{\mathrm{dl}}}{K}\mathbf{I}_{K}$; $\mathbf{F}_{\textrm{BB}}= \left [ \mathbf{f}_{1}^{\textrm{BB}},\mathbf{f}_{2}^{\textrm{BB}},\cdots ,\mathbf{f}_{K}^{\textrm{BB}} \right ]\in \mathbb{C}^{K\times K}$ and $\mathbf{F}_{\textrm{RF}}= \left [ \mathbf{f}_{1}^{\textrm{RF}},\mathbf{f}_{2}^{\textrm{RF}},\cdots ,\mathbf{f}_{K}^{\textrm{RF}} \right ]\in \mathbb{C}^{N_{\textrm{BS}}\times K}$ represent the digital and analog TPCs, respectively. Moreover, $n_{k}^{\mathrm{dl}}\in \mathbb{C}$ denotes the noise at the $k$-th user, which follows a complex Gaussian distribution with zero mean and variance $\sigma_{\mathrm{dl}}^{2} $. Since  $\mathbf{F}_{\textrm{RF}}$ is implemented by phase shifters, the modulus of its elements is constant and normalized to satisfy $\left|\left [ \mathbf{F}_{\textrm{RF}} \right ] _{m,n}\right|^{2}=\frac{1}{N_{\textrm{BS}}}$. Furthermore, the elements in $\mathbf{F}_{\textrm{RF}}$ have the general form $\left [ \mathbf{F}_{\textrm{RF}} \right ]_{m,n}=\frac{1}{N_{\textrm{BS}}}e^{j\phi _{m,n}}$, where $\phi _{m,n}$ is the quantized phase\cite{sun,hybrid_precoding_AA_1}. The digital TPC $\mathbf{F}_{\textrm{BB}}$ has to satisfy the power normalization constraint $\quad\quad\left\|\mathbf{F}_{\textrm{RF}}\mathbf{f}_{k}^{\textrm{BB}}\right\|_\textrm{F}^2=1, 
k=1,2, \ldots, K .$ to ensure that the digital TPC does not provide any power gain.

Furthermore, a block-fading channel is assumed\cite{hybrid_precoding_AA_1,hybrid_precoding_AA_2}, and the achievable rate of the $k$-th user can be calculated as 
\begin{equation}\label{sum_rate}
	\setlength\abovedisplayskip{3pt}
	\setlength\belowdisplayskip{3pt}
	\begin{split}
		R_{k}= \textrm{log}_{2}\left ( 1+\frac{\frac{P_{\textrm{dl}}}{K}\left| \mathbf{h}_{k}^{\textrm{dl}}\mathbf{F}_{\textrm{RF}}\mathbf{f}_{k}^{\textrm{BB}}\right|^{2}}{\frac{P_{\textrm{dl}}}{K}\sum _{n\neq k}\left|\mathbf{h}_{k}^{\textrm{dl}}\mathbf{F}_{\textrm{RF}}\mathbf{f}_{n}^{\textrm{BB}} \right|^{2}+\sigma_{\mathrm{dl}}^{2}} \right ).
	\end{split}
\end{equation}
\vspace{-0.5cm}
\subsection{Near-Field Channel Model}

In the near-field region, the spherical wave model needs to be adopted\cite{cui_1,han_yu_1,xlmimo}. Accordingly, the near-field channel is represented as
\begin{equation}\label{near-h}
	\setlength\abovedisplayskip{3pt}
	\setlength\belowdisplayskip{3pt}
	\begin{split}
		\textbf{h}^{\textrm{near}}_{k}=\sqrt{N_{\textrm{BS}}}\alpha _{0}\mathbf{b}\left ( r _{0} ,\phi _{0}\right )+\sqrt{\frac{N_{\textrm{BS}}}{L-1}}\sum_{l=1}^{L-1}\alpha _{l}\mathbf{b}\left ( r _{l} ,\phi _{l}\right ),
	\end{split}
\end{equation}
where $r_{0}$ denotes the distance from the center of the ULA of the BS to the user and $r_{l}$ represents the distance from the center of the ULA of the BS to the scatterer of the $l$-th path. Moreover, a single LoS path and $L-1$ NLoS paths are considered. Specifically, the near-field array steering vector $\mathbf{b}\left ( r_{l},\phi_{l}  \right )$ can be expressed as
\begin{equation}\label{near-b}
	\setlength\abovedisplayskip{3pt}
	\setlength\belowdisplayskip{3pt}
	\begin{split}
		\mathbf{b}\left ( r_{l},\phi_{l}  \right )=\frac{1}{\sqrt{N_{\textrm{BS}}}}\left [  e^{j\frac{2\pi }{\lambda}\left ( r_{l}^{(0)}-r_{l} \right )},..., e^{j\frac{2\pi }{\lambda}\left ( r_{l}^{(N_{\textrm{BS}}-1)}-r_{l} \right )} \right ]^{\textrm{T}},
	\end{split}
\end{equation}
where $r_{l}^{(n)}$ denotes the distance from the $n$-th antenna at the BS to the user $(l=0)$ or the $l$-th scatterer $(l\ge 1)$ and $r_{l}^{(n)}=\sqrt{r_{l}^{2}+\delta _{n}^{2}d^{2}-2r_{l}\phi _{l} \delta _{n}d    } $,  where $\delta _{n} =\frac{2n-N_{\mathrm{BS} } +1}{2} ,n=0,1,\dots ,N_{\mathrm{BS} }-1$.

\vspace{-0.5cm}
\subsection{Problem Formulation}
\vspace{-0.1cm}
Our objective is to maximize the sum rate of the system, which can be expressed as $R_{\textrm{sum}}=\sum_{k=1}^{K}R_{k}$ based on (\ref{sum_rate}). It can be seen from (\ref{sum_rate}) that analog TPC $\mathbf{F}_{\textrm{RF}}$ and digital TPC $\mathbf{F}_{\textrm{BB}}$ have to be designed jointly for maximizing the sum rate of the system.

The analog TPC $\mathbf{F}_{\textrm{RF}}$ comprises $K$ analog beamforming vectors, which will generate $K$ beams pointing at the $K$ users in the DL to generate beamforming gains\cite{hybrid_precoding_AA_1,hybrid_precoding_AA_2}. Due to the hardware limitations of the RF chains, the beamforming vectors can typically take only finite deterministic values. Accordingly, some finite-size predefined codebooks were devised to solve this problem\cite{codebook_for_hp}, and $K$ codewords are selected from the codebook to form the analog TPC. 

In the far-field domain, the discrete Fourier transform (DFT) codebook based on the far-field array steering vector is widely adopted\cite{qi_1}, which contains $N_{\textrm{BS}}$ codewords corresponding to the $N_{\textrm{BS}}$ angles sampled at equal intervals. The beam formed by the codewords in conventional DFT codebooks is also called narrow beams.


Additionally, also in the far-field domain, a wide beam codebook having lower resolution was proposed for reducing the resources required to search for the optimal codeword. The $n$-th codeword in the wide beam codebook is given by
\begin{equation}\label{wide_codeword}
	\setlength\abovedisplayskip{3pt}
	\setlength\belowdisplayskip{3pt}
	\begin{split}
		\mathbf{e}\left ( \varphi _{n}^{\textrm{w}} \right )=\sqrt{\frac{M}{N_{\textrm{BS}}}}\left [ 1,e^{j\frac{2\pi d}{\lambda }\textrm{sin}\varphi _{n}^{\textrm{w}} } ,...,e^{j\frac{2\pi d}{\lambda }\left ( \frac{N_{\textrm{BS}}}{M} -1 \right )\textrm{sin}\varphi _{n}^{\textrm{w}} }\right ]^{\textrm{T}},
	\end{split}
\end{equation}
where $\varphi _{n}^{\textrm{w}}=\textrm{arcsin}\left ( -1+\frac{2n-1}{N_{\textrm{BS}}/M} \right )$, $n=1,2,\cdots,N_{\textrm{BS}}/M$, denotes the $n$-th sampling angle for the wide beam; $M$ represents the number of narrow beams that can be covered by each wide beam. 
The far-field wide beam codebook can be expressed as 
\begin{equation}\label{far_wide_codebook}
	\setlength\abovedisplayskip{3pt}
	\setlength\belowdisplayskip{3pt}
	\begin{split}
		\mathcal{W}=\left\{\mathbf{e}\left( \varphi _{1}^{\textrm{w}}\right), \ldots, \mathbf{e}\left(\varphi _{n}^{\textrm{w}}\right), \ldots,\right. \left.\mathbf{e}\left( \varphi _{N_{\textrm{BS}}}^{\textrm{w}}\right)\right\}^{\ast }.
	\end{split}
\end{equation}

Observe from (\ref{wide_codeword}) that the dimension of the wide-beam codeword is only $N_{\textrm{BS}}/M$, which means that its implementation requires the activation of only $N_{\textrm{BS}}/M$ antennas at the BS. 

In general, the codebooks designed for the far-field domain is not applicable to channels in the near-field domain. When employing beams formed by far-field codewords for users in the near-field domain, the beams will not be accurately aligned and significant energy leakage will occur\cite{cui_1,han_yu_1}.
\begin{figure}[t]
	\centering
	\includegraphics[width=2.5in]{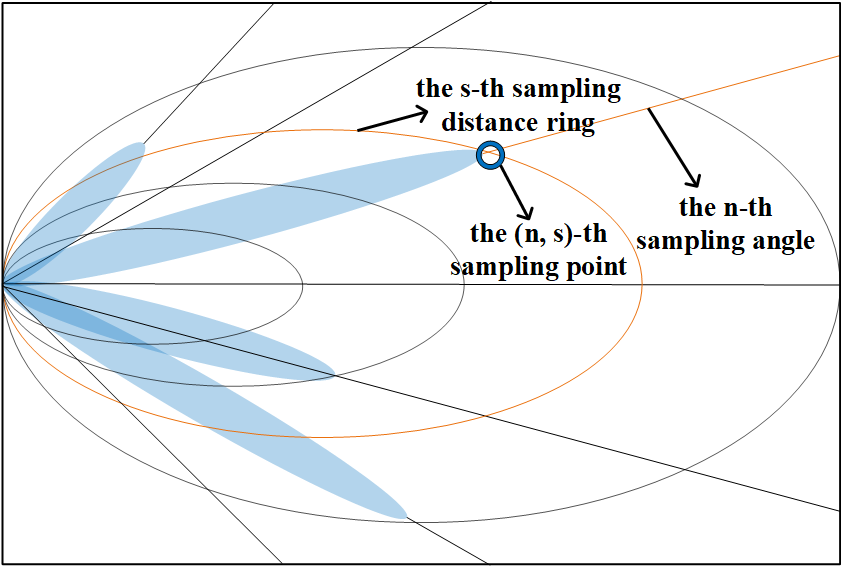}
	\vspace{-0.2cm}
	\caption{Near-field codebook with angle sampling and distance sampling. }
	\label{near_codebook_fig}
	\vspace{-0.5cm}
\end{figure}

Hence, for near-field channels, a near-field codebook is adopted, which additionally incorporates distance based sampling and thus contains several times more codewords than the far-field codebook.
Specifically, the near-field codebook uniformly divides the space into $N_{\textrm{BS}}$ angles in terms of direction, while dividing the entire space into $S$ distance rings in terms of distance, which is illustrated in Fig. \ref{near_codebook_fig}.
Let us define the distance from the BS to the intersection of the $s$-th distance ring and the $n$-th direction, i.e., the $[(s-1)N_{\textrm{BS}}+n]$-th sampling point, as $r_{n}^{s}$. Then, the near-field codebook can be expressed as
\begin{equation}\label{near_codebook}
	\setlength\abovedisplayskip{3pt}
	\setlength\belowdisplayskip{3pt}
	\begin{split}
			\mathcal{N}=\left\{\mathbf{b} \left ( \varphi _{1},r_{1}^{1}\right ),\dots ,\mathbf{b} \left ( \varphi _{N_{\textrm{BS}}},r_{N_{\textrm{BS}}}^{1}\right ),\dots ,\mathbf{b} \left ( \varphi _{N_{\textrm{BS}}},r_{N_{\textrm{BS}}}^{S}\right )\right\}.
	\end{split}
\end{equation}
For representational convenience, the index of the  $[(s-1)N_{\textrm{BS}}+n]$-th near-field codeword can also be denoted by $(n, s)$, where $s$ is the distance index and $n$ is the direction index.

By adopting the near-field codebook, the problem of maximizing the sum rate can be formulated as
\begin{equation}\label{problem_1}
	\setlength\abovedisplayskip{3pt}
	\setlength\belowdisplayskip{3pt}
	\begin{split}
	& \max _{\mathbf{F}_{\textrm{RF}}, \mathbf{F}_{\textrm{BB}}} \sum_{k=1}^K \log _2\left(1+\frac{\frac{P_{\mathrm{dl}}}{K}\left| \mathbf{h}_k^{\textrm{dl}} \mathbf{F}_{\textrm{RF}} \mathbf{f}_k^{\textrm{BB}}\right|^2}{\frac{P_{\textrm{dl}}}{K} \sum_{i \neq k}\left| \mathbf{h}_k^{\textrm{dl}} \mathbf{F}_{\textrm{RF}} \mathbf{f}_i^{\textrm{BB}}\right|^2+\sigma_{\mathrm{dl}}^2}\right) \\
	& \text { s.t. }\left[\mathbf{F}_{\textrm{RF}}\right]_{:, k}= \mathbf{f}_k^{\textrm{RF}}\in \mathcal{N}, \quad k=1,2, \ldots, K ,\\
	& \quad\quad\left\|\mathbf{F}_{\textrm{RF}}\mathbf{f}_{k}^{\textrm{BB}}\right\|_\textrm{F}^2=1, 
	\quad k=1,2, \ldots, K .
\end{split}
\end{equation}

Note that the globally optimal solution cannot be obtained by existing algorithms because Problem (\ref{problem_1}) is non-convex. Based on existing literature \cite{hybrid_precoding_AA_1,sun}, a common technique of solving Problem (\ref{problem_1}) is to design $\mathbf{F}_{\textrm{RF}}$ given a fixed $\mathbf{F}_{\textrm{BB}}$, and then design $\mathbf{F}_{\textrm{BB}}$ by relying on either the zero-forcing (ZF) or the minimum mean-squared error (MMSE) criterion. Specifically, the BS first selects the optimal $K$ codewords that maximize the beam gain for each user, which is the same as beam training. The corresponding problem can be formulated as
\begin{equation}\label{problem_2}
	\setlength\abovedisplayskip{3pt}
	\setlength\belowdisplayskip{3pt}
	\begin{split}
		& \max _{\mathbf{f}_{k}^{\textrm{RF}}}  \left| \mathbf{h}_{k}^{\textrm{dl}}\mathbf{f}_{k}^{\textrm{RF}}\right|,\quad k=1,2, \ldots, K , \\
		& \text { s.t. }\left[\mathbf{F}_{\textrm{RF}}\right]_{:, k}= \mathbf{f}_k^{\textrm{RF}}\in \mathcal{N}, \quad k=1,2, \ldots, K, \\
	\end{split}
\end{equation}
where $g_{k}=\left| \mathbf{h}_{k}^{\textrm{dl}}\mathbf{f}_{k}^{\textrm{RF}}\right|$ denotes the beamforming gain generated by $\mathbf{f}_{k}^{\textrm{RF}}$ for the $k$-th user. After obtaining $\mathbf{F}_{\textrm{RF}}$, the effective channel $\mathbf{h}_{k}^{\textrm{dl}}\mathbf{F}_{\textrm{RF}}$ can be estimated at the BS relying on the uplink (UL) pilot signals. Finally, given the estimated effective channel, the digital TPC $\mathbf{F}_{\textrm{BB}}$ is designed based on either the ZF or the MMSE criterion. More details are provided in the next section.

\section{Three-phase GNN-based Beam Training}\label{DL_model}

In this section, we describe the proposed GNN-based beam training scheme, which comprises three phases, i.e., estimation based on GNN, near-field beam allocation, and design of the hybrid TPC.


\vspace{-0.3cm}
\subsection{UL Pilot Transmission}

We propose to only test the far-field wide beams during UL pilot transmission, because the corresponding test information was shown to be sufficient for neural networks to determine the optimal near-field beam \cite{my_letter,self_2,jiangguoli}. Since the size of the far-field wide-beam codebook is much smaller than that of the near-field codebook, the UL pilot overhead is significantly reduced. 
In this subsection, we firstly detail the transmission of the UL pilot signals and then introduce the structure of the GNN.

In TDD communication systems, the UL channel and the DL channel are reciprocal, i.e., $ \mathbf{h}_{k}^{\textrm{ul}}=\left ( \mathbf{h}_{k}^{\textrm{dl}} \right )^{\textrm{T}}$, where $\mathbf{h}_{k}^{\textrm{ul}}$ denotes the UL channel of the $k$-th user. During UL pilot transmission, the $K$ users are assumed to send mutually orthogonal pilot signals. To this end, orthogonal sequences are employed as pilot signals. Let us define the pilot signal sent by the $k$-th user by $ \sqrt{P_{\textrm{ul}}}\mathbf{x}_{k}\in \mathbb{C}^{1\times K}$, where $\mathbf{x}_{k}$ satisfies $\mathbf{x}_{i}\mathbf{x}_{j}^{\mathrm{H}}=0$ if $i\neq j$ and $\mathbf{x}_{i}\mathbf{x}_{i}^{\mathrm{H}}=1$, and $P_{\textrm{ul}}$ is the UL pilot power. Since $N_{\textrm{RF}}$ RF chains are employed at the BS, the BS may select $N_{\textrm{RF}}$ different codewords from the far-field wide beam codebook to construct analog TPC for combining the signals received from the users, and then the beamforming gain of the selected $N_{\textrm{RF}}$ codewords for each user can be derived. Note that the digital TPC is set to $\mathbf{I}_{K}$ during UL pilot transmission. 


Since the far-field wide beam codebook is adopted, only $N_{\textrm{BS}}/M$ antennas are activated at the BS and the corresponding dimension is reduced from $N_{\textrm{BS}}$ to $N_{\textrm{BS}}/M$. For example, $\mathbf{h}_{k}^{\textrm{w,ul}}\in \mathbb{C}^{\left ( N_{\textrm{BS}}/M \right )\times 1 }$ and $\mathbf{F}_{\textrm{RF}}^{\textrm{w}}\in \mathbb{C}^{\left ( N_{\textrm{BS}}/M \right ) \times K}$ represent the DL channel and analog TPC, respectively, when far-field wide-beam codewords are employed. The superscript ``$\textrm{w}$'' indicates that the far-field wide beam codebook is adopted. Therefore, the signal received by the BS during the $t$-th UL pilot transmission can be expressed as
\begin{equation}\label{receive_1}
	\setlength\abovedisplayskip{3pt}
	\setlength\belowdisplayskip{3pt}
	\begin{aligned}
		\mathbf{Y}^{\textrm{w},(t)}=\sum_{k=1}^{K}\sqrt{P_{\textrm{ul}}}\left (  \mathbf{F}_{\textrm{RF}}^{\textrm{w},(t)}\right ) ^{\mathrm{T} } \mathbf{h}_{k}^{\textrm{w},\textrm{ul}}\mathbf{x}_{k}+\left (  \mathbf{F}_{\textrm{RF}}^{\textrm{w},(t)}\right ) ^{\mathrm{T} } \mathbf{N}^{\textrm{w,ul}},
	\end{aligned}
\end{equation}
where $\mathbf{N}^{\textrm{w,ul}}\in \mathbb{C}^{\left ( N_{\textrm{BS}} /M\right ) \times K}$ denotes the noise at the BS during UL transmission and each column of $\mathbf{N}^{\textrm{w,ul}}$ is independently distributed as $\mathcal{C N}\left(\mathbf{0}, \sigma_{\textrm{ul}}^2 \mathbf{I}\right)$. $\mathbf{F}_{\textrm{RF}}^{\textrm{w,(t)}}=\left [\mathbf{e} _{\left ( t,1 \right )},\mathbf{e} _{\left ( t,2 \right )},\dots ,\mathbf{e} _{\left ( t,N_{\textrm{RF}} \right )}\right ]$ represents the analog TPC consisting of the $N_{\textrm{RF}}$ far-field wide beam codewords selected during the $t$-th UL pilot transmission. $\mathbf{e} _{\left ( t,n \right )}=\mathbf{e}\left ( \varphi _{(t-1)n}^{\textrm{w} }\right )$, $ t=1,2,\dots,T, n=1,2,\dots, N_{\textrm{RF}}$, denotes the $n$-th codeword selected by the BS during the $t$-th UL pilot transmission. By exploiting the orthogonality of the pilot signals, the transmitted signals of the $k$-th user can be obtained by left-multiplying with $\frac{\mathbf{x}_{k}^{\ast}}{\sqrt{P_{\textrm{ul}}}}$, which leads to 
\vspace{-0.2cm}
\begin{equation}\label{receive_4}
	\setlength\abovedisplayskip{4pt}
	\setlength\belowdisplayskip{3pt}
	\begin{aligned}
		\mathbf{r}_{k}^{\textrm{w,ul}(t)} &= \frac{\mathbf{x}_{k}^{\ast }}{\sqrt{P_{\textrm{ul}}}}\left ( \sum_{k=1}^{K}\sqrt{P_{\textrm{ul}}}  \mathbf{F}_{\textrm{RF}}^{\textrm{w},(t)\mathrm{T}} \mathbf{h}_{k}^{\textrm{w},\textrm{ul}}\mathbf{x}_{k}+ \mathbf{F}_{\textrm{RF}}^{\textrm{w},(t)\mathrm{T}} \mathbf{N}^{\textrm{w,ul}}\right )^{\textrm{T}}\\
		&= \mathbf{x}_{k}^{\ast }\mathbf{x}_{k}^{\textrm{T}}\left ( \mathbf{h}_{k}^{\textrm{w},\textrm{ul}} \right )^{\textrm{T}}\mathbf{F}_{\textrm{RF}}^{\textrm{w},(t)}+\frac{\mathbf{x}_{k}^{\ast }\left ( \mathbf{N}^{\textrm{w,ul}} \right )^{\textrm{T}}\mathbf{F}_{\textrm{RF}}^{\textrm{w},(t)}}{\sqrt{P_{\textrm{ul}}}}\\
		&=\left [ \mathbf{h}_{k}^{\textrm{w},\textrm{dl}}\mathbf{e} _{\left ( t,1 \right )},\mathbf{h}_{k}^{\textrm{w},\textrm{dl}}\mathbf{e} _{\left ( t,2 \right )},\dots ,\mathbf{h}_{k}^{\textrm{w},\textrm{dl}}\mathbf{e} _{\left ( t,N_{\textrm{RF}} \right )}\right ]\\&+\frac{\mathbf{x}_{k}^{\ast }\left ( \mathbf{N}^{\textrm{w,ul}} \right )^{\textrm{T}}\mathbf{F}_{\textrm{RF}}^{\textrm{w,(t)}}}{\sqrt{P_{\textrm{ul}}}},
	\end{aligned}
\end{equation}

Apparently, the beamforming gain of the $(t, n)$-th far-field wide beam codeword for the $k$-th user can be roughly estimated as $\hat{g}_{\left ( t,n \right )} = \left| \left [ \mathbf{r}_{k}^{\textrm{ul}(t)} \right ]_{n}\right|$.

 Accordingly, after $T=\frac{N_{\textrm{BS}}/M}{N_{\textrm{RF}}}$ transmissions, the far-field wide-beam gain information matrix for the $k$-th user can be obtained and is denoted as
\vspace{-0.2cm}
\begin{equation}\label{receive_5}
	\begin{aligned}
	\mathbf{R}_{k}^{\textrm{w}}=\left [\left(\mathbf{r}_{k}^{\textrm{w,ul}(1)}\right)^{\textrm{T}},\left(\mathbf{r}_{k}^{\textrm{w,ul}(2)}\right)^{\textrm{T}},\dots , \left(\mathbf{r}_{k}^{\textrm{w,ul}(T)}\right)^{\textrm{T}} \right ]^{\textrm{T}}.
	\end{aligned}
\end{equation}

\begin{figure*}[t]
	\centering
	\includegraphics[width=6in]{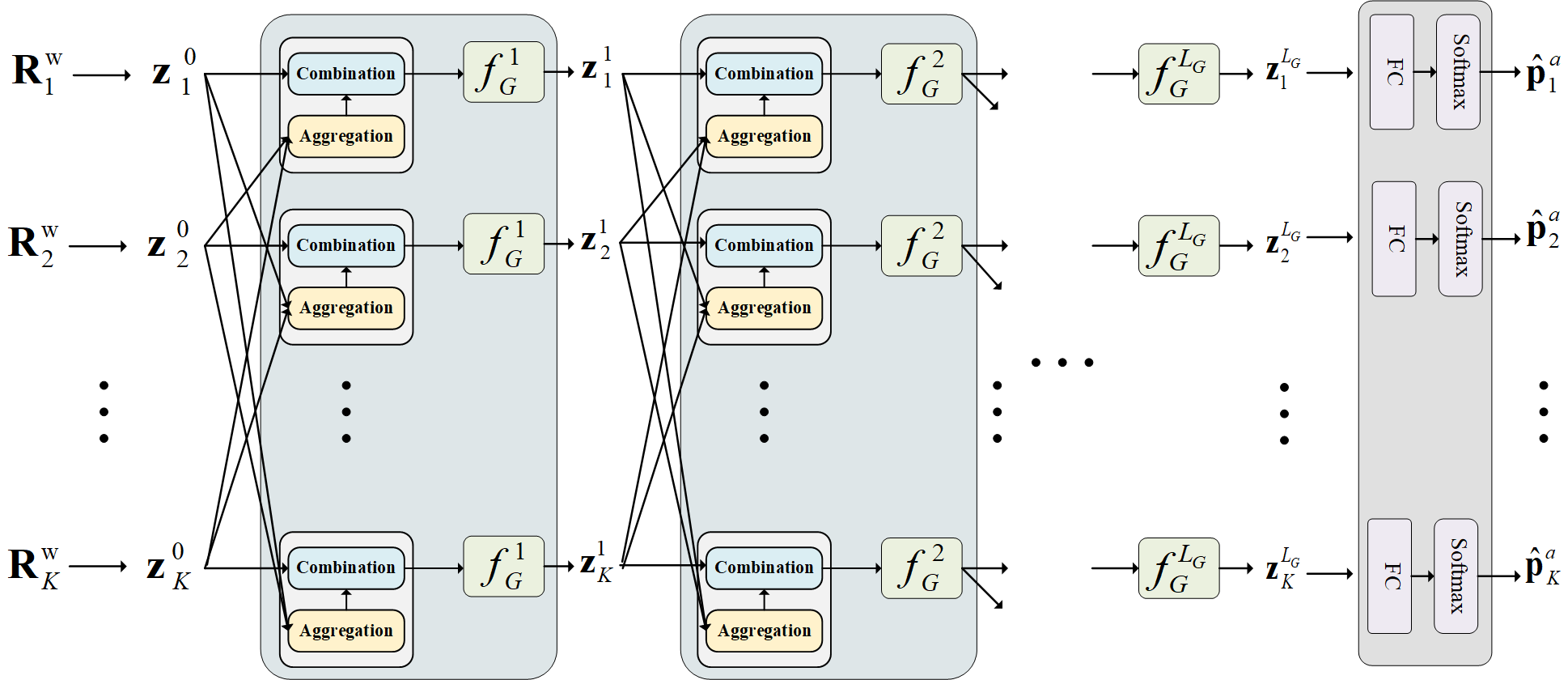}
	\vspace{-0.2cm}
	\caption{Overall architecture of GNN-based estimation network.}
	\label{gnn}
	\vspace{-0.6cm}
\end{figure*}

However, the optimal near-field codeword for each user cannot be derived directly from $\left\{ \mathbf{R}_{k}^{\textrm{w}}\right\}_{k=1}^{K}$, yet $\left\{ \mathbf{R}_{k}^{\textrm{w}}\right\}_{k=1}^{K}$ implicitly contains information about a user's optimal near-field codeword\cite{my_letter,jiangguoli}. Therefore, we propose to employ a neural network to construct a mapping from the beamforming gain information obtained from the far-field wide beams to the optimal near-field codeword. It should be noted that conventional estimation methods can not be readily employed for this mapping because of the complex nonlinear relationship between $\left\{ \mathbf{R}_{k}^{\textrm{w}}\right\}_{k=1}^{K}$ and the optimal near-field codeword for each user. Mathematically, this UL beam training model can be formulated as 
\begin{equation}\label{mapping}
	\begin{aligned}
	\left\{\mathbf{b}_{1}^{\star }, \mathbf{b}_{2}^{\star },\cdots ,\mathbf{b}_{K}^{\star }\right\}=f_{\textrm{nn}}\left ( \mathbf{R}_{1}^{\textrm{w}},\mathbf{R}_{2}^{\textrm{w}} ,\cdots ,\mathbf{R}_{K}^{\textrm{w}}\right ),
	\end{aligned}
\end{equation}
where $\mathbf{b}_{k}^{\star }\in \mathcal{N}$ denotes the optimal near-field codeword of the $k$-th user, while  $f_{\textrm{nn}}\left ( \cdot  \right )$ represents the mapping function of the neural network.

To achieve high estimation accuracy for the optimal near-field codeword, GNNs are employed for constructing this mapping by learning. Compared to traditional algorithms and ordinary FCNN or CNN, GNNs have the following unrivalled advantages\cite{jiang_tao_1}:
\begin{enumerate}
	\item Powerful ability to learn and fit non-linear relationships;
	\item Capability to exploit the relationships between users and improve the estimation performance through information gleaned from the surrounding users;
	\item Adaptability to any number of users;
\end{enumerate}

The above advantages are general properties of GNN and are discussed in detail in the remainder of this section. Our proposed scheme is the first to exploit these properties to improve the performance and reduce the overhead of near-field beam training.

\vspace{-0.3cm}
\subsection{Architecture of GNN-based Estimation Network}
We first introduce the architecture of the proposed GNN-based estimation network, which comprises processing, feature updating, and output modules.

$1)$ $\textit{Preprocessing Module}$: The main function of the preprocessing module is to transform the input of the GNN, i.e., the far-field wide-beam beamforming gain matrix of each user into real-valued vectors that allow it to be recognized and further processed by the neural network. Specifically, let us define the initial feature vector of the $k$-th user as $\mathbf{z}_{k}^{0}$ given by
\vspace{-0.1cm}
\begin{equation}\label{input}
	\setlength\belowdisplayskip{3pt}
	\begin{aligned}
		\mathbf{z}_{k}^{0}=\left [ \textrm{vec}\left ( \mathfrak{R}\left ( \mathrm{R}_{k}^{\textrm{w}} \right ) \right )^{\textrm{T}},\textrm{vec}\left ( \mathfrak{I}\left ( \mathrm{R}_{k}^{\textrm{w}} \right ) \right )^{\textrm{T}} \right ]^{\textrm{T}},
	\end{aligned}
\end{equation}
where $\textrm{vec}\left ( \cdot  \right )$ stands for the vectorization operation, $\mathfrak{R} \left ( \cdot  \right ) $ and $\mathfrak{I} \left ( \cdot  \right ) $ denote the real and imaginary parts of a complex number, respectively. The initial feature vectors are further processed by the feature updating module. 

$2)$ $\textit{Graph-based Feature Updating Module}$: The feature updating module mainly comprises $L_{\textrm{G}}$ feature updating layers. Let us define the mapping function of the $l$-th feature updating layer as $f^{l}_{\textrm{G}}\left ( \cdot  \right )$ and the feature vector of the $k$-th user updated by the $l$-th feature updating layer as $\mathbf{z}_{k}^{l}$. In fact, the feature updating layers have the same structure and functionality as a conventional fully-connected layer, and their mapping function can be formulated as:
\vspace{-0.2cm}
\begin{equation}\label{fc_func}
	\begin{aligned}
		 f^{l}_{\textrm{G}}\left ( \mathbf{v}_{l} \right )=\textrm{ReLU}\left ( \mathbf{W}_{l} \mathbf{v}_{l}+ \mathbf{bias}_{l}\right ),
	\end{aligned}
\vspace{-0.2cm}
\end{equation}
where $ \mathbf{W}_{l}$ and $\mathbf{bias}_{l}$ denote the weight matrix and bias vector of the $l$-th updating layer, respectively; $\textrm{ReLU}\left ( \cdot  \right )$ denotes the nonlinear activation function, which is widely applied for deep learning-based classification tasks;  $\mathbf{v}_{l}$ denotes the input of the $l$-th updating layer. In addition, the number of neurons in each feature update layer is the same and is set to $V$.

Unlike conventional neural networks, the feature updating action of GNN is based on a graph structure, which is illustrated in Fig. \ref{gnn}. Specifically, feature vector updating at each layer also requires the aggregation and the combination of the feature vectors of neighbouring users. In the GNN, this updating process can be expressed as
\begin{equation}\label{gnn_func}
	\begin{aligned}
	\mathbf{z}_{k}^{l}=f^{l}_{\textrm{G}}\left ( f_{\textrm{combine}}\left ( \mathbf{z}_{k}^{l-1} ,f_{\textrm{aggregate}}\left ( \left\{ \mathbf{z}_{j}^{l-1}\right\} _{j\in \mathcal{O}_{(k)}}\right )\right )\right ),
	\end{aligned}
\end{equation}
where $\mathcal{O}_{(k)}=\left \{ 1,2,\dots ,k-1,k+1,\dots ,K \right \} $ represents the neighbouring users of the $k$-th user, and $f_{\textrm{aggregate}}\left ( \cdot  \right )$ and $f_{\textrm{combine}}\left ( \cdot  \right )$ represent the aggregation and combination functions, respectively. Given this structure,  the feature updating action exploits not only the feature vectors of user $k$ itself, but also those of the neighbouring users.

In 6G usage scenarios, especially in massive communication scenarios, the user-density is high \cite{ITU}. Hence, the adjacent users would have similar wireless propagation environments as well as similar system  parameters such as noise intensity, path loss, scatterers and reflection coefficients. Additionally, they tend to have similar LoS and NLoS paths. The wireless propagation environment and system parameters affect the pilot signals, in other words, the pilot signals would contain the information of the wireless propagation environment and system parameters. As a result, the pilot signals of the adjacent users can be exploited as ``copies", which will assist the neural network in better distinguishing the LoS and NLoS paths, identifying scatterers, and mitigating the impact of the noise. 

\begin{figure*}[t]
	\centering
	\includegraphics[width=6in]{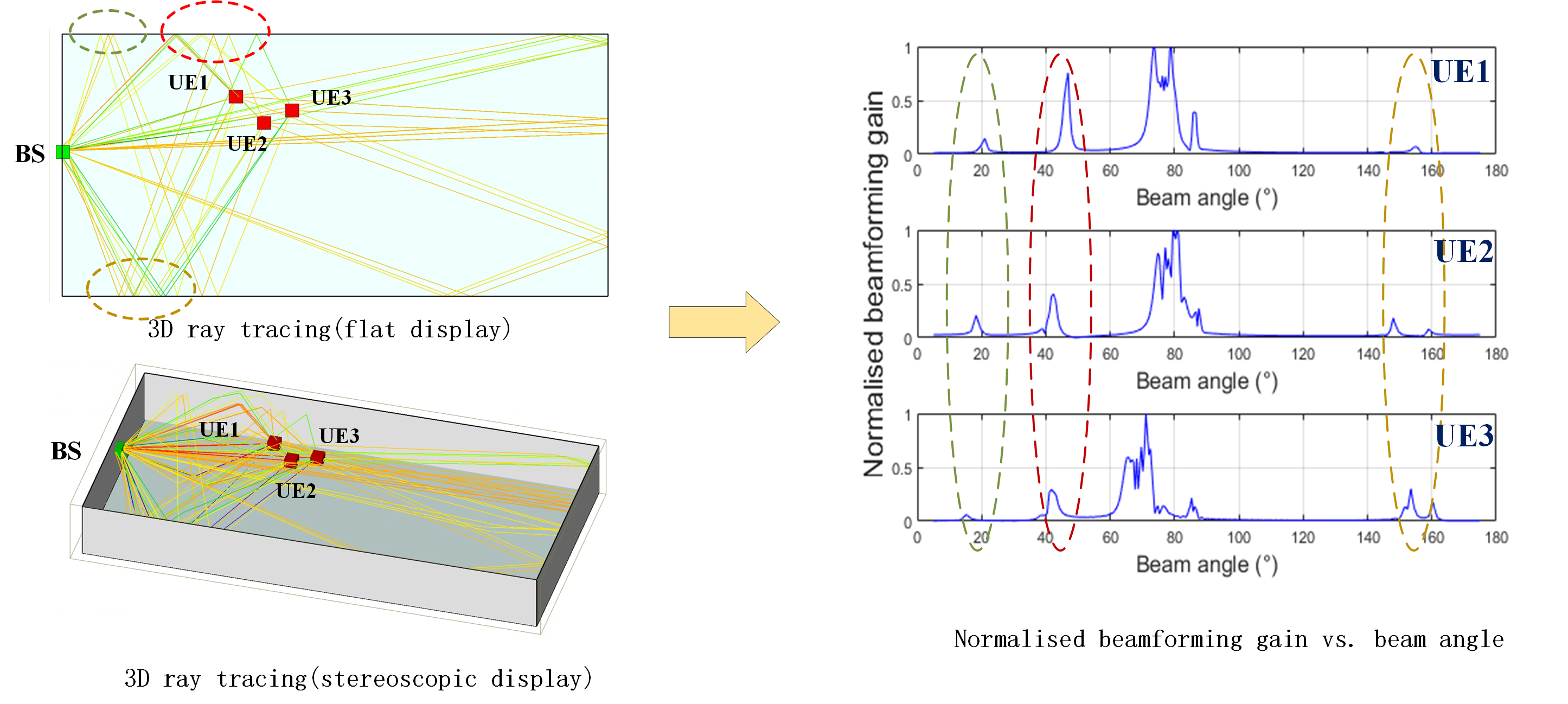}
	\vspace{-0.2cm}
	\caption{Neighbouring users' beamforming gain information is mutually correlated, as they share the wireless propagation environment.}
	\label{environment}
	\vspace{-0.6cm}
\end{figure*}

To illustrate this more intuitively, we simulate the wireless communication environment of a smart factory by using the ray tracing software Wireless Insite\cite{WI}, as shown in Fig. \ref{environment}, where each ray represents a propagation path. Furthermore, the three users' UL beamforming gain information, which is obtained based on the pilot signals, is shown in Fig. \ref{environment}. Observe from Fig. \ref{environment} that the similarity of the wireless propagation environments, system parameters and paths of the different users results in similar received pilot signals and similar beamforming gain information. We note that the above simulations do not account for noise. When considering a noisy environment, it can be expected that the beamforming gain information of other users can assist the neural network in better discriminating between noise-induced anomalies and the true LoS path, which can be regarded as a form of ``diversity gain".

Consequently, the design of the aggregation function $f_{\textrm{aggregate}}\left ( \cdot  \right )$ and the combination function $f_{\textrm{combine}}\left ( \cdot  \right )$ is crucial for the performance of the GNN. An effective aggregation function based on GNN theory\cite{jiang_tao_1} can take the following form:
\vspace{-0.2cm}
\begin{equation}\label{agg_func}
	\begin{aligned}
		\left [ f_{\textrm{aggregate}}\left ( \left\{ \mathbf{z}_{j}^{l-1}\right\} _{j\in \mathcal{O}_{(k)}}\right ) \right ]_{i}
		&=\textrm{mean}\left ( \left [ \mathbf{z}_{1}^{l-1} \right ]_{i},\cdots ,\left [ \mathbf{z}_{K}^{l-1} \right ]_{i} \right )\\&=
		\left [ \tilde{\mathbf{z}}_{k}^{l-1} \right ]_{i},
	\end{aligned}
\end{equation}
where $\textrm{mean}\left ( \cdot  \right )$ represents the operation of element-wise averaging, and $\tilde{\mathbf{z}}_{k}^{l-1}$ denotes the aggregated feature vector.  Element-wise averaging has the important advantage of constant dimensionality, which allows GNNs to accommodate any arbitrary number of users. Specifically, regardless of how many users there are, the dimensionality of the aggregated feature vectors $\left [ \tilde{\mathbf{z}}_{k}^{l-1} \right ]_{i}$ remains invariant, because the average pooling is element-wise. This property implies that the number of neurons in each layer of the network is independent of the number of users. By contrast, this is hard to achieve in ordinary FCNN and CNN, if they are to exploit the information of other users. 

As for the combination function, we adopt the concatenation operation, which leads to
\vspace{-0.2cm}
\begin{equation}\label{com_func}
	\begin{aligned}
		f_{\textrm{combime}}\left ( \mathbf{z}_{k}^{l-1} ,\tilde{\mathbf{z}}_{k}^{l-1}\right )=\left [ \left ( \mathbf{z}_{k}^{l-1} \right )^{\textrm{T}} ,\left (\tilde{\mathbf{z}}_{k}^{l-1}  \right )^{\textrm{T}}\right ]^{\textrm{T}}.
	\end{aligned}
\end{equation}

$3)$ $\textit{Output Module}$: The output of the final layer of the feature updating module, $\mathbf{z}_{k}^{L_{\mathbf{G}}}$, is fed to the output module for final processing. Since the number of near-field codewords is finite, GNN-based UL beam training based on Problem (\ref{problem_2}) can be regarded as a classification problem. Therefore, a pair of fully connected layers and a Softmax activation layer are employed to construct the output module. The probability vector is output after the Softmax layer, which contains the probability of each near-field codeword becoming the optimal codeword estimated by the neural network. 

$4)$ $\textit{Distance and Angle Network}$: We note that since the number of near-field codewords is very large, it is difficult to train the neural network, when employing only one GNN network to estimate the optimal near-field codeword for each user, hence leading to slow convergence and underfitting problems. In view of this, we propose a dual neural network structure based on the properties of near-field codebooks. Specifically, we design and train a pair of GNN-based estimation networks, i.e., a distance network and an angle network, which estimate the distance index and angle index of the optimal near-field codeword, respectively. The structures of the input module and the feature updating module of the two networks are exactly the same, only the output dimension of the fully connected layer of the output module is different and corresponds respectively to the total number of angle and distance indices. Based on the dual neural network structure, the same UL far-field wide-beam beamforming gain information is fed into both networks and the pair of probability vectors of each user are represented as
\vspace{-0.2cm}
\begin{equation}\label{prob_vector}
	\begin{gathered}
		\hat{\mathbf{p}}_{k}^{\textrm{a}}=\left [ \hat{p}_{k,1}^{\textrm{a}},\hat{p}_{k,2}^{\textrm{a}},\cdots ,\hat{p}_{k,N_{\textrm{BS}}}^{\textrm{a}}\right ]^{T},\\
		\hat{\mathbf{p}}_{k}^{\textrm{d}}=\left [ \hat{p}_{k,1}^{\textrm{d}},\hat{p}_{k,2}^{\textrm{d}},\cdots ,\hat{p}_{k,S}^{\textrm{d}}\right ]^{T},
	\end{gathered}
\end{equation}
where $\hat{p}_{k,s}^{\textrm{d}}$ and $\hat{p}_{k,n}^{\textrm{a}}$ denote the probability that user $k$'s optimal near-field codeword has the distance indexed by $s$ and the angle indexed by $n$, respectively. Based on $\hat{\mathbf{p}}_{k}^{\textrm{a}}$ and $\hat{\mathbf{p}}_{k}^{\textrm{d}}$, the probability that the $(n,s)$-th near-field codeword, i.e., $\mathbf{b} \left ( \varphi _{n},r_{n}^{s}\right )$, is the optimal near-field codeword for user $k$ is given by
\vspace{-0.2cm}
\begin{equation}\label{prob_code}
	\begin{aligned}
	\hat{p}_{k}^{(n,s)}=\hat{p}_{k,n}^{\textrm{a}}\hat{p}_{k,s}^{\textrm{d}}.
	\end{aligned}
\end{equation}

Thus, the probability vector of user $k$ with respect to all near-field codewords can be represented as
\vspace{-0.1cm}
\begin{equation}\label{prob_vector_code}
	\begin{aligned}
		\hat{\mathbf{p} }_{k}=\left [ \hat{p}_{k}^{\textrm{(1,1)}},\hat{p}_{k}^{\textrm{(2,1)}},\cdots,\hat{p}_{k}^{(N_{\mathrm{BS}},S)} \right ],
	\end{aligned}
\vspace{-0.2cm}
\end{equation}
where a larger $\hat{p}_{k}^{(n,s)} $ means that the optimal codeword of the $k$-th user is more likely to be $\mathbf{b} \left ( \varphi _{n},r_{n}^{s}\right )$.

\begin{figure*}[t]
	\centering
	\includegraphics[width=6.2in]{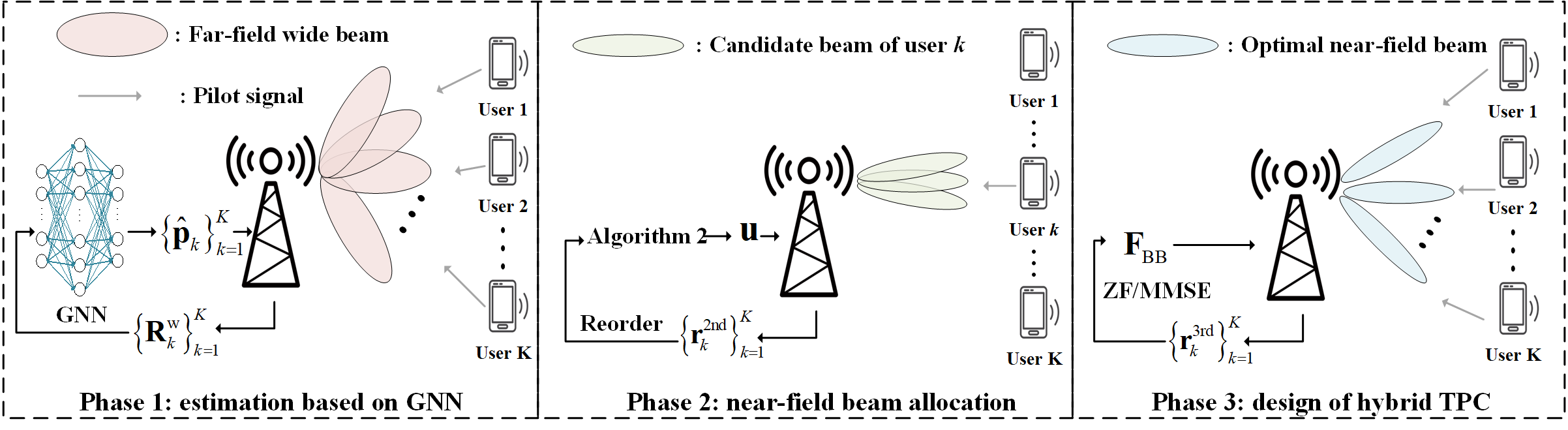}
	\vspace{-0.2cm}
	\caption{Proposed three-phase GNN-based beam training.}
	\label{phase}
	\vspace{-0.6cm}
\end{figure*}

\vspace{-0.3cm}
\subsection{Three-Phase GNN-based Beam training}

Given the structure of the GNN-based estimation networks, we describe the proposed three-phase GNN-based multi-user near-field beam training and hybrid TPC design, which is presented in Algorithm \ref{alg1}. The proposed three-phase GNN-based beam training is illustrated in Fig. \ref{phase}. For the proposed scheme, pilots are required three times. They are used for the estimation of the optimal near-field beam, the beam assignment, and the estimation of the effective channel, respectively. Overall, the number of pilot symbols required by our proposed scheme is only $N_{\mathrm{BS} }/M+K+K$, which is much lower than the number of pilot symbols required to test all near-field codewords. 
\begin{algorithm}[t]
	\caption{ Three-Phase GNN-based Beam training Scheme}
	\label{alg1}
	\begin{algorithmic}[1] 
		\REQUIRE ~~\\ 
		Trained GNN-based distance network and angle network;  Far-field wide-beam codebook $\mathcal{W} $; Near-field codebook $\mathcal{N} $; \\
		\ENSURE ~~\\ 
		Analog TPC $\mathbf{F}_{\textrm{RF}}$ and digital TPC $\mathbf{F}_{\textrm{BB}}$;
		\STATE Obtain all users' far-field wide-beam gain information $\left\{ \mathbf{R}_{k}^{\textrm{w}}\right\}_{k=1}^{K}$ via (\ref{receive_4}) and (\ref{receive_5});
		\STATE Obtain the probability vectors of each user's near-field codewords $\left \{ \hat{\mathbf{p} }_{k} \right \} _{k=1}^{K} $ by feeding $\left\{ \mathbf{R}_{k}^{\textrm{w}}\right\}_{k=1}^{K}$ into the angle and distance networks, respectively;
		\STATE Apply Algorithm 2 to assign a beam to each user and obtain the index vector $\mathbf{u}$;
		\STATE Obtain the analog TPC $\mathbf{\tilde{F} }_{\textrm{RF}}$ according to  $\mathbf{u}$ via (\ref{designed_frf});
		\STATE Estimate the effective channel $	\mathbf{\hat{\mathbf{H} } } _{\mathrm{ef} } $ via (\ref{h_hat_2}) and obtain the digital precoder $\tilde{\mathbf{F}}_{\textrm{BB}}$ based on ZF or MMSE criterion via (\ref{designed_fbb}).
		\RETURN $\mathbf{\tilde{F} }_{\textrm{RF}}$, $\tilde{\mathbf{F}}_{\textrm{BB}}$ 
	\end{algorithmic}

\end{algorithm}

$1)$ $\textit{Estimation Phase Based on GNN}$: For the proposed scheme, the $K$ users in the system firstly transmit orthogonal pilot sequences of length $K$ in the UL, and the BS combines these signals by employing $\mathbf{F}_{\textrm{RF}}$ consisting of $N_{\textrm{RF}}$ different far-field wide-beam codewords. The signals of the $k$-th user received at the BS, i.e., $\mathbf{r}_{k}^{\textrm{w,ul}(t)}$,  are given by (\ref{receive_4}). After  $T=\frac{N_{\textrm{BS}}/M}{N_{\textrm{RF}}}$ transmissions, the $k$-th user's beamforming gain information about all far-field wide-beam codewords, i.e., gain information matrix $\mathbf{R}_{k}^{\textrm{w}}$, is obtained in (\ref{receive_5}).

Subsequently, we input the beamforming gain information matrices $\left\{ \mathbf{R}_{k}^{\textrm{w}}\right\}_{k=1}^{K}$ of all users into the trained angle estimation and distance estimation networks, respectively. Accordingly, the probability  vector of each user regarding all near-field codewords, i.e. $\left \{ \hat{\mathbf{p} }_{k} \right \} _{k=1}^{K} $, is given by (\ref{prob_vector}) and (\ref{prob_vector_code}).


$2)$ $\textit{Near-field Beam Allocation Phase}$: In this phase, we first test the candidate codewords of each user, which allows us to resolve beam conflicts, thereby improving the sum rate of the system. We define the near-field codewords corresponding to the $K$ largest probabilities in $\hat{\mathbf{p} }_{k}$ as the candidate codewords of user $k$, and the set of their indices is denoted as $\mathcal{L} _{k} $, i.e.,
\begin{equation}\label{set_l}
	\begin{aligned}
	\left \{ \hat{p}_{k}^{(\sigma _{k}^{1},\gamma _{k}^{1} ) } ,\hat{p}_{k}^{(\sigma _{k}^{2},\gamma _{k}^{2} ) },\dots ,\hat{p}_{k}^{(\sigma _{k}^{N_{\mathrm{BS}}S},\gamma _{k}^{N_{\mathrm{BS}}S} ) }\right \}\\=\left \langle \left \{\hat{p}_{k}^{\textrm{(1,1)}},\dots , \hat{p}_{k}^{(N_{\mathrm{BS}},S)}\right \}  \right \rangle  ,
	\end{aligned}
\end{equation}
and
\begin{equation}\label{set_l}
	\begin{aligned}
			\mathcal{L} _{k}  =\left \{ (\sigma _{k}^{1},\gamma _{k}^{1}) ,(\sigma _{k}^{2},\gamma _{k}^{2}),\dots ,(\sigma _{k}^{K},\gamma _{k}^{K})\right \},
	\end{aligned}
\end{equation}
where $\left \langle \cdot  \right \rangle$ denotes the  ordering operation, e.g., for $\mathcal{A}=\left\{a_{1}, a_{2}, \ldots, a_{n}\right\},\langle\mathcal{A}\rangle=\left\{a_{\sigma_{1}}, a_{\sigma_{2}}, \ldots, a_{\sigma_{n}}\right\}$ with $a_{\sigma_{1}} \geq a_{\sigma_{2}} \geq \ldots \geq a_{\sigma_{n}}$. 

After obtaining $\left \{ \mathcal{L} _{k}  \right \} _{k=1}^{K} $, all users transmit the UL pilot signals again to test the candidate codewords. Since the candidate codewords of each user are not necessarily identical, we use time-orthogonal pilot signals during this pilot transmission period, such that the users sequentially transmit a pilot symbol $x_{k}$. In the $k$-th time slot, the BS receives the pilot symbol from user $k$ and employs the corresponding near-field codewords to combine it. The received signal is given by 
\begin{equation}\label{rece_test}
	\begin{aligned}
		\mathbf{r}_{k}^{\textrm{2nd}}&= \left ( \sqrt{P_{\textrm{ul}}}\mathbf{F}_{\textrm{RF}}^{(k)\textrm{T}}\mathbf{h}_{k}^{\textrm{ul}}x_{k}+\mathbf{F}_{\textrm{RF}}^{(k)\textrm{T}}\mathbf{n}^{\textrm{ul}}
		\right )^{\textrm{T}}\\
		&=\sqrt{P_{\textrm{ul}}}\left [\mathbf{h} _{k}^{\mathrm{dl} } \mathbf{b} \left ( \varphi _{\sigma _{k}^{1}},r_{\sigma _{k}^{1}}^{\gamma _{k}^{1}}\right ),\dots ,\mathbf{h} _{k}^{\mathrm{dl} } \mathbf{b} \left ( \varphi _{\sigma _{k}^{K}},r_{\sigma _{k}^{K}}^{\gamma _{k}^{K}}\right ) \right ]\\&+\mathbf{n}^{\textrm{ul}}\mathbf{F}_{\textrm{RF}}^{(k)},
		\end{aligned}
\end{equation}
where $\mathbf{F}_{\textrm{RF}}^{(k)}=\left [ \mathbf{b} \left ( \varphi _{\sigma _{k}^{1}},r_{\sigma _{k}^{1}}^{\gamma _{k}^{1}}\right ),\dots , \mathbf{b} \left ( \varphi _{\sigma _{k}^{K}},r_{\sigma _{k}^{K}}^{\gamma _{k}^{K}}\right ) \right ]$ denotes the analog TPC consisting of the candidate codewords of user $k$, $\mathbf{n}^{\textrm{ul}}\in \mathbb{C} ^{N_{\mathrm{BS} }\times 1 } $ denotes the noise, and pilot symbol $x_{k}$ is assumed to be 1. 

After $K$ time slots, the BS has obtained the beamforming gain information of the candidate codewords for all users, i.e., $\left \{ 	\mathbf{r}_{k}^{\textrm{2nd}} \right\} _{k=1}^{K}$. Note that only $K$ time slots are required for this pilot transmission, which is equal to all user transmitting an orthogonal sequence of length $K$ simultaneously in the previous phase.

Based on the gain information $\left \{ 	\mathbf{r}_{k}^{\textrm{2nd}} \right\} _{k=1}^{K}$, we reorder the candidate codewords for each user based on the modulus of the corresponding received signal. For the $k$-th user, the reordering process is formulated as 
\begin{equation}\label{r_sec_order}
	\begin{aligned}
		\left \{ \left | \left [ \mathbf{r}_{k}^{\textrm{2nd}} \right ]_{\eta _{1} }   \right | ,\left | \left [ \mathbf{r}_{k}^{\textrm{2nd}} \right ]_{\eta _{2} }   \right |,\dots ,\left | \left [ \mathbf{r}_{k}^{\textrm{2nd}} \right ]_{\eta _{K} }   \right | \right \} =\\
		\left \langle \left \{ \left | \left [ \mathbf{r}_{k}^{\textrm{2nd}} \right ]_{1 }   \right | ,\left | \left [ \mathbf{r}_{k}^{\textrm{2nd}} \right ]_{2 }   \right |,\dots ,\left | \left [ \mathbf{r}_{k}^{\textrm{2nd}} \right ]_{K }   \right | \right \} \right \rangle .
	\end{aligned}
\end{equation}
We then define $\mathbf{r}_{k}^{\textrm{sort}}=\left [  \left | \left [ \mathbf{r}_{k}^{\textrm{2nd}} \right ]_{\eta _{1} }   \right | ,\left | \left [ \mathbf{r}_{k}^{\textrm{2nd}} \right ]_{\eta _{2} }   \right |,\dots ,\left | \left [ \mathbf{r}_{k}^{\textrm{2nd}} \right ]_{\eta _{K} }   \right | \right ] 
$ as the sorted modulus vector, and define the index of the codewords corresponding to $\left [ \mathbf{r}_{k}^{\textrm{sort}} \right ] _{l}$  as $c_{k} (l)$. Finally, we obtain the sorted index vector $\mathbf{c} _{k} =\left [ c_{k} (1),c_{k} (2),\dots ,c_{k} (K) \right ] $.

After obtaining the sorted index vectors $\left \{\mathbf{c} _{k}\right \}_{k=1}^{K}$ and the sorted modulus vectors $\left \{\mathbf{r}_{k}^{\textrm{sort}}\right \}_{k=1}^{K}$ of all users, we have to assign appropriate beams to each user such that beam conflicts are avoided. Specifically, when two users are close to each other, their optimal beams may coincide. However, when the same beam is employed to serve two different users, there will be serious interference between the users regardless of how the digital TPC is designed. 
Therefore, we cannot directly assign the codeword corresponding to $c_{k} (1)$ to user $k$, as it may cause serious beam conflict. To overcome this problem, we propose a beam allocation scheme based on $\left \{\mathbf{c} _{k}\right \}_{k=1}^{K}$ and $\left \{\mathbf{r}_{k}^{\textrm{sort}}\right \}_{k=1}^{K}$ to resolve beam conflicts in the near field. The proposed beam allocation scheme is outlined in Algorithm 2. 

The main idea of this beam assignment scheme is to assign beams to each user based on the signal modulus of the beams in $\left \{\mathbf{r}_{k}^{\textrm{sort}}\right \}_{k=1}^{K}$. When the optimal beams of user $i$ and user $j$ are identical, i.e., $c_{i} (1)=c_{j} (1)$, the beam is assigned to user $i$ if $\left [ \mathbf{r}_{i}^{\textrm{sort}} \right ] _{1} $ is greater than $\left [ \mathbf{r}_{j}^{\textrm{sort}} \right ] _{1} $ and vice versa, which is conducive to achieving a high sum rate.

To this end, firstly, we initialize a $K$-dimensional index vector $\mathbf{u}$ to store the index of each user's last assigned near-field codeword. Next, in Step 4, we search for the highest priority user $k_{max}$, which has the currently highest modulus of the pilot signal. If the beam corresponding to $c_{k_{max}}(1)$ has not yet been assigned, then the beam will be assigned to user $k_{max}$ and the set $\mathcal{R} $ will be updated, which corresponds to Steps 5 and 6. If the codeword corresponding to $c_{k_{max}}(1)$ has already been assigned, then user $k_{max}$ has to give up this beam and delete index  $c_{k_{max}}(1)$ from vector $\mathbf{c} _{k}$. User $k_{max}$ continues to join the next round of beam assignment. We repeat Steps 4 to 8 until set $\mathcal{R}$ becomes empty, i.e., all users are assigned beams. Finally, we obtain index vector $\mathbf{u}$, where $\left [ \mathbf{u} \right ] _{k} $ represents the index of the codeword that user $k$ is finally assigned. For convenience, we denote the angle index and distance index of the codeword corresponding to $\left [ \mathbf{u} \right ] _{k} $ as $\tilde{n} _{k} $ and $\tilde{s} _{k} $, respectively, where $\left [ \mathbf{u} \right ] _{k}=\left ( \tilde{s} _{k}-1 \right )N_{\mathrm{BS} } + \tilde{n} _{k} $.

 \begin{algorithm}[t]
	\caption{Probability vector-based beam allocation algorithm.}
	\label{alg2}
	\begin{algorithmic}[1] 
		\REQUIRE ~~\\ 
		$\left \{ \hat{\mathbf{p} }_{k} \right \} _{k=1}^{K} $\\
		\ENSURE ~~\\ 
		Index vector $\mathbf{u}$; 
		\STATE Get the index of the candidate codeword for each user $\left \{ \mathcal{L} _{k}  \right \} _{k=1}^{K} $ and the corresponding modal value of the pilot signal $\left \{\mathbf{r}_{k}^{\mathrm{2nd}}\right \}_{k=1}^{K}$ via (\ref{set_l}) and (\ref{rece_test});
		\STATE Sort the candidate codewords to obtain the sorted index vectors $\left \{\mathbf{c} _{k}\right \}_{k=1}^{K}$ and the sorted modulus vectors $\left \{\mathbf{r}_{k}^{\textrm{sort}}\right \}_{k=1}^{K}$ of all users via (\ref{r_sec_order});
		\STATE Initialization: $\mathbf{u} \gets \mathbf{0}_{K}$,  $\mathcal{R} =\left \{ 1,2,\dots ,K \right \} $
		\WHILE{ $\mathcal{R} \ne \emptyset$}
		\STATE Determine the highest-priority users $k_{max} =\mathop{\arg\max}\limits_{k\in \mathcal{R} }\left [ \mathbf{r}_{k}^{\textrm{sort}} \right ]_{1}  $;
		\IF{$\forall i,\left [ \mathbf{c} _{k_{max} } \right ]_{1} \ne \left [ \mathbf{u} \right ] _{i} $ }
		\STATE $\left [  \mathbf{u}  \right ] _{k_{max} } \gets \left [ \mathbf{c} _{k_{max} } \right ]_{1}$ ;
		\STATE  $\mathcal{R} \gets \mathcal{R}\setminus \left \{ k_{max}  \right \}  $;
		\ELSE
		\STATE $\mathbf{c} _{k_{max} } \gets \left [ \mathbf{c} _{k_{max} } \right ] _{2:\mathrm{end} } $;
		\STATE $\mathbf{r}_{k_{max} }^{\textrm{sort}}\gets \left [ \mathbf{r}_{k_{max} }^{\textrm{sort}} \right ]_{2:\mathrm{end} } $; 
		\ENDIF
		\ENDWHILE 
		\RETURN $\mathbf{u}$; 
	\end{algorithmic}
\end{algorithm}

$3)$ $\textit{Design of Hybrid TPC}$: At this stage, we present the design of the analog TPC $\mathbf{F} _{\mathrm{RF} } $ and digital TPC $\mathbf{F} _{\mathrm{BB} } $. Based on the index vector $\mathbf{u}$ obtained with Algorithm 2, the analog TPC is given by
\vspace{-0.2cm}
\begin{equation}\label{designed_frf}
  \begin{aligned}
	\mathbf{\tilde{f} }_k^{\textrm{RF}}=\mathbf{b} \left ( \varphi _{\tilde{n} _{k}},r_{\tilde{n} _{k}}^{\tilde{s} _{k}}\right ),
	\mathbf{\tilde{F} }_{\textrm{RF}}=\left [  \mathbf{\tilde{f} }_1^{\textrm{RF}}, \mathbf{\tilde{f} }_2^{\textrm{RF}} ,\dots , \mathbf{\tilde{f} }_K^{\textrm{RF}}\right ].
  \end{aligned}
\vspace{-0.2cm}
\end{equation} 
	
After obtaining the analog TPC, all users transmit an orthogonal pilot sequence of length $K$, when the analog TPC is set as $\mathbf{\tilde{F} }_{\textrm{RF}}$  and the digital TPC is set as $\mathbf{I} _{K}$. According to (\ref{receive_1}) and (\ref{receive_4}), the  received signal of the $k$-th user is given by 
\vspace{-0.2cm}
\begin{equation}\label{receive_6}
	\begin{aligned}
		\mathbf{r} _{k}^{\mathrm{3rd} }=\left [ \mathbf{h}_{k}^{\mathrm{dl} } \mathbf{\tilde{f} }_1^{\textrm{RF}} ,\mathbf{h}_{k}^{\mathrm{dl} } \mathbf{\tilde{f} }_2^{\textrm{RF}} ,\dots ,\mathbf{h}_{k}^{\mathrm{dl} } \mathbf{\tilde{f} }_K^{\textrm{RF}}   \right ]+\frac{\mathbf{x}_{k}^{\ast }\left ( \mathbf{N}^{\textrm{ul}} \right )^{\textrm{T}}\mathbf{\tilde{F} }_{\textrm{RF}}}{\sqrt{P_{\textrm{ul}}}}  , 
	\end{aligned}
\vspace{-0.2cm}
\end{equation}
where $\mathbf{r} _{k}^{\mathrm{3rd} }$ denotes the received pilot signal in hybrid TPC design phase. Subsequently, the effective channel can be estimated as\cite{sun,hybrid_precoding_AA_1,hybrid_precoding_AA_2}
\begin{equation}\label{h_hat_2}
	\begin{aligned}
		\mathbf{\hat{\mathbf{H} } } _{\mathrm{ef} } =\left [\left (  \mathbf{r} _{1}^{\mathrm{3rd}  }  \right ) ^{\mathrm{T} },\left (  \mathbf{r} _{2}^{\mathrm{3rd}  }  \right ) ^{\mathrm{T} },\dots ,\left (  \mathbf{r} _{K}^{\mathrm{3rd}  }  \right ) ^{\mathrm{T} }  \right ] ^{\mathrm{T} }.
	\end{aligned}
\end{equation}
Finally, we design the digital TPC by using the ZF or the MMSE criterion, respectively, which leads to
\vspace{-0.2cm}
\begin{equation}\label{designed_fbb}
	\begin{aligned}
		\tilde{\mathbf{F}}_{\textrm{BB}}^{\textrm{ZF}}&=\hat{\mathbf{H}}_{\textrm{ef}}^{\textrm{H}}\left ( \hat{\mathbf{H}}_{\textrm{ef}}\hat{\mathbf{H}}_{\textrm{ef}}^{\textrm{H}} \right )^{-1},\\
		\tilde{\mathbf{F}}_{\textrm{BB}}^{\textrm{MMSE}}&=\hat{\mathbf{H}}_{\textrm{ef}}^{\textrm{H}}\left ( \frac{P_{\textrm{dl}}}{K}\hat{\mathbf{H}}_{\textrm{ef}}\hat{\mathbf{H}}_{\textrm{ef}}^{\textrm{H}} +\sigma _{\textrm{dl}}^{2}\mathbf{I}_{K}\right )^{-1}.
	\end{aligned}
\end{equation}

\subsection{Computational Complexity}

In this subsection, we quantify the computational complexity of the proposed three-phase GNN-based beam training scheme, which is mainly determined by the forward propagation of the GNNs and the design of the digital TPCs. In the feature updating module of the dedicated angle and distance networks, the computational complexity for $K$ users is on the order of $\mathcal{O} \left (K V N_{\mathrm{BS} } /M +KV^{2}  L_{\mathrm{G} }  \right )  $, where $ L_{\mathrm{G} }$ is the number of feature updating layers and $V$ is the number of neurons in each feature updating layer. In the output module of the angle and distance networks, the computational complexity of the $K$ users is $\mathcal{O} \left ( KV N_{\mathrm{BS} } +KVS \right ) $. In the design of the digital TPC, the computational complexity is $\mathcal{O} \left (K^{3} \right ) $, which is mainly imposed by the ZF or MMSE algorithms. Therefore, the overall computational complexity is as follows:
\begin{equation}\label{compute}
	\begin{aligned}
\mathcal{O} \left (K V N_{\mathrm{BS} } /M +KV^{2}  L_{\mathrm{G} }  +KV N_{\mathrm{BS} } +KVS +K^{3}\right ).
	\end{aligned}
\end{equation}

\vspace{-0.4cm}

\section{Simulation Results}\label{simulation}


\vspace{-0.2cm}
\subsection{System Setup}

Unlike other ordinary neural network models, the training and performance of a GNN relies on real environmental information, which requires us to generate realistic and accurate wireless channel samples. Accordingly, the state-of-the-art ray-tracing algorithm-based software called Wireless Insite \cite{WI} is adopted to simulate realistic wireless environments and generate accurate channel data. 

\begin{figure}[t]
	\centering
	\includegraphics[width=3in]{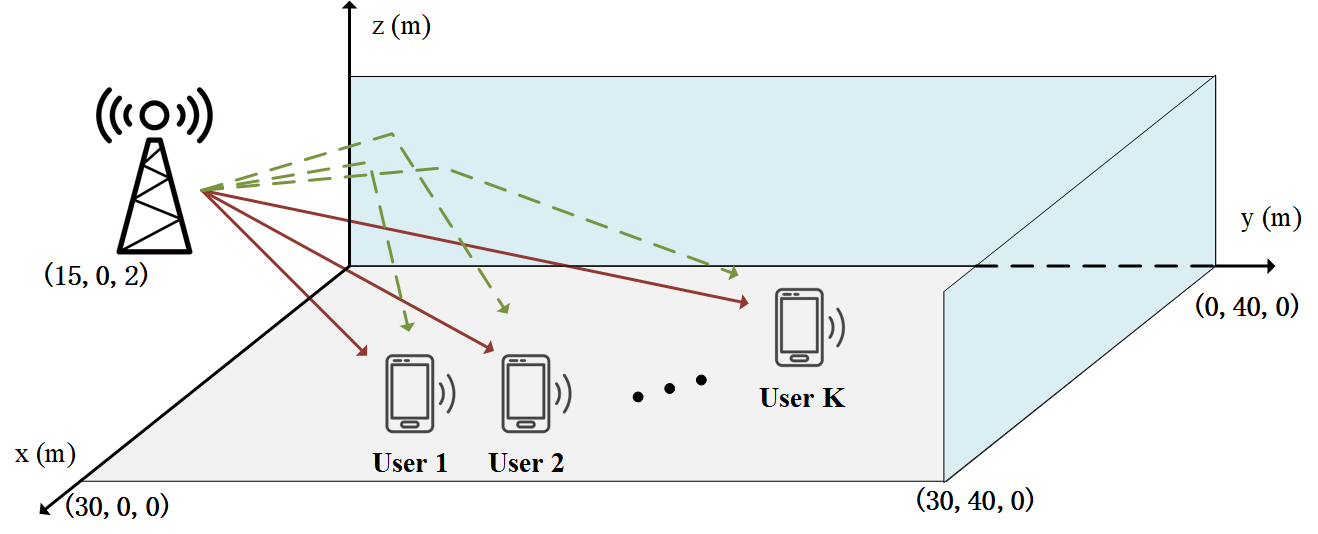}
	\vspace{-0.2cm}
	\caption{Simulation layout for  multiuser XL-MIMO system.}
	\label{factory}
	\vspace{-0.4cm}
\end{figure}

Specifically, we utilized the WI software to create a wireless communication scenario resembling a large smart factory as shown in Fig. 5, where the surrounding walls, ceilings, and floors serve as the main sources of scatterers. In this scenario, we consider a millimetre wave communication system, where a BS equipped with $N_{\mathrm{BS} } =256$ antennas serves $K=8$ single-antenna users. As shown in Fig. \ref{factory}, the users are randomly distributed in the factory and the BS is fixed at the location with coordinates (15, 0, 2) meters. For the high user density, the distance between any two users is less than 6 meters. The length, width, and height of the factory building are 40 metres, 30 metres, and 5 metres, respectively. In addition, the range of the user's coordinates $(x, y, z)$ is set to $x\in (0\ \mathrm{m} ,40\ \mathrm{m} )$, $y\in(0\ \mathrm{m} ,30\ \mathrm{m} )$, and $z = 1\ \mathrm{m} $. Hence, the users are guaranteed to be in the near-field domain. 

For the simulation of the wireless communication environment, we use a sinusoidal waveform with effective bandwidth $B=10 \mathrm{~MHz} $ and centre carrier frequency $f_{c} = 30 \mathrm{~GHz}$ as the waveform of the transmitted signals. Both the BS and the users employ half-wave dipole antennas and the spacing of the antennas is set to $d=\frac{\lambda }{2} $. The transmit power of the UL pilot signal and the DL data signal are set to $4\mathrm{~dBm}$ and $5\mathrm{~dBm}$, respectively, and the noise power is set to $-81\mathrm{~dBm}$ for both UL and DL, unless stated otherwise. 
Based on the given system parameters, the number of angle and distance samples of the near-field codebook are set to $N_{\mathrm{BS} } =256$ and $S=5$, respectively. Therefore, the total number of codewords in the near-field codebook is $N_{\mathrm{BS} }S=1280$. The wide beam in the far-field wide-beam codebook is set to cover 4 narrow beams, i.e., $M = 4$, such that the total number of codewords in the far-field wide-beam codebook is equal to $N_{\mathrm{BS} }/M=64$.

Furthermore, we used a three-layer GNN, i.e., $L_{\textrm{G}}$=3, for the feature updating module. The specific parameters of the neural network structure are summarized in Table \ref{netwrok_table}. For the last fully connected layer, the output size of the angle network was 256, while that of the distance network was 5.
We generated 12000 channel samples based on the given system parameters and calculated their corresponding labels, where 90\% of the data was utilized to train the proposed GNN and the remaining 10\% was utilized to evaluate the performance of the GNN. In the course of training, the Adam optimizer was employed.
The learning rate decay strategy was also applied along with an initial learning rate of 0.006. The learning rate decayed by half when there was no significant improvement in the estimation accuracy within two epochs. Furthermore, the proposed GNN was trained for 50 epochs, where the size of the batch for each epoch was 800.

\begin{table}[t]
		\centering 
	\caption{\text { Network Parameters }} 
	\label{netwrok_table} 
	\resizebox{0.95\columnwidth}{!}{
	\begin{tabular}{|c|c|c|}
		\hline
		\textbf{Module Name}                                 & \textbf{Network Layer Name} & \textbf{(Input Size, Output Size)} \\ \hline
		\multirow{3}{*}{Graph-based Feature Updating Module} & GNN(ReLU)                   & (128,128)                          \\ \cline{2-3} 
		& GNN(ReLU)                   & (128,128)                          \\ \cline{2-3} 
		& GNN(ReLU)                   & (128,128)                          \\ \hline
		\multirow{2}{*}{Output Module}                       & FC(ReLU)                    & (128,128)                          \\ \cline{2-3} 
		& FC(Softmax)                 & (128,5 or 256)                     \\ \hline
	\end{tabular}}
\vspace{-0.5cm}
\end{table}
\vspace{-0.4cm}
\subsection{Metrics and Benchmarks}

We consider three performance metrics to evaluate our proposed scheme, namely the sum rate, effective sum rate, and estimation accuracy.

1) The sum rate $R_{\mathrm{sum} }$ is given by $R_{\mathrm{sum} }=\sum_{k=1}^{K} R_{k} $,
where $R_{k}$ denotes the achievable rate of the $k$-th user, which is defined in (\ref{sum_rate}).

2) The effective sum rate is given by 
\begin{equation}\label{criteria_2}
	\setlength\abovedisplayskip{3pt}
	\setlength\belowdisplayskip{3pt}
	R_{\mathrm{eff}}=\left ( 1-\frac{T_{\mathrm{p}} }{T_{\mathrm{t}} }  \right ) R_{\mathrm{sum}} ,
\end{equation}
where $T_{\mathrm{t}}$ represents the total time of a communication session and $T_{\mathrm{p}}$ denotes the time consumed by pilot signal transmission in a communication session \cite{make1}. In general, $T_{\mathrm{p}}$ is the product of the number of UL pilot symbols and the time required to transmit one pilot symbol. In the simulation, we set the time required to transmit a pilot symbol, i.e., a time slot, to 0.1 $\mu$s and the total time of a communication session $T_{\mathrm{t}}$  to 0.2 ms, respectively. The effective sum rate accounts for both the sum rate and the pilot overhead, and only schemes that can achieve a high sum rate at a low pilot overhead provide a desirable effective sum rate.

3) The estimation accuracy of the adopted neural networks is defined as $A_{\mathrm{cc}}=\frac{K_{\mathrm{r}}}{K} $. Recall from (\ref{set_l}) that $ (\sigma _{k}^{1},\gamma _{k}^{1})$ is the index of the codeword corresponding to the maximum probability value in $\hat{\mathbf{p} }_{k}$. If $ (\sigma _{k}^{1},\gamma _{k}^{1})$ is also the index of the optimal near-field codeword of user $k$ without considering beam conflicts, then the estimate of the neural network with respect to user $k$ is correct and otherwise, it is incorrect. $K_{r}$ denotes the number of users for which the neural network provides the correct estimate.
 

In order to better characterize the performance of the proposed scheme, four benchmark schemes are considered:

1) $Exhaustive$ $\mathit{near}$-$field$ $search$:The exhaustive near-field scheme tests all near-field codewords during UL pilot transmission, instead of testing the far-field codewords. Thus, it acquires beamforming gain information about all near-field codewords for each user. In addition, the beam allocation scheme outlined in Algorithm \ref{alg2} is also employed to avoid beam conflicts. The exhaustive search is expected to achieve the best performance but causes excessive pilot overhead.

2) $FC$ $neural$ $network$-$based$ $scheme$ \cite{qi_1}: In the FC neural network-based scheme, FC neural networks are employed for separately estimating optimal near-field beam of each user and the analog, as well as digital TPC, is designed based on the optimal beam for each user.

3) $CNN$-$based$ $scheme$\cite{make1}: Similar to the FC neural network-based scheme, we apply CNNs to separately estimate the optimal near-field beam for each user, and the corresponding analog and digital TPCs are designed based on the obtained near-field beams.

4) $OMP$-$based$ $scheme$: The OMP algorithm is a classical compressive sensing (CS) based channel estimation algorithm, which leverages the sparsity of millimetre wave channels for determining the resulting low-dimensional channel with low pilot overhead \cite{cs_1,cui_1}. In the OMP-based scheme, we first estimate the channel of each user using the OMP algorithm and obtain the estimated value of the channel $\left \{ \mathbf{\hat{h} } _{k}^{\mathrm{dl} }  \right \} _{k=1}^{K} $. Then, the principle of maximum ratio combining (MRC) is employed and the analog TPC is designed as 
\begin{equation}\label{omp_rate}
	\mathbf{\tilde{F} _{\mathrm{RF} } } =\left [ \frac{\left ( \mathbf{\hat{h} } _{1}^{\mathrm{dl} } \right )^{\mathrm{H} }}{\left \| \mathbf{\hat{h} } _{1}^{\mathrm{dl} } \right \| _{2}  }  , \frac{\left ( \mathbf{\hat{h} } _{2}^{\mathrm{dl} } \right )^{\mathrm{H} }}{\left \| \mathbf{\hat{h} } _{2}^{\mathrm{dl} } \right \| _{2}  }, \dots ,\frac{\left ( \mathbf{\hat{h} } _{K}^{\mathrm{dl} } \right )^{\mathrm{H} }}{\left \| \mathbf{\hat{h} } _{K}^{\mathrm{dl} } \right \| _{2}  }\right ] .
\end{equation}
Finally, the effective channel is estimated by transmitting additional pilots in the UL and the corresponding digital TPC is designed.

\vspace{-0.3cm}
\subsection{Simulation Results}

\begin{figure*}
	\begin{minipage}[t]{0.49\linewidth}
		\centering
		\includegraphics[width=3.2in]{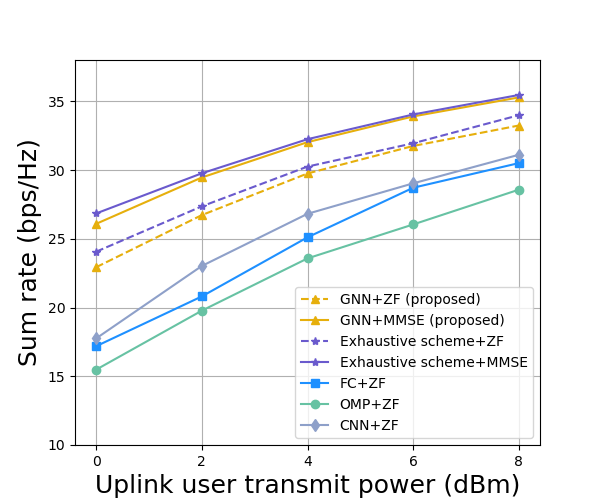}
		\vspace{-0.2cm}
		\caption{Sum rate of different schemes as function of the uplink user transmit power. }
		\label{up_sumrate}\vspace{-0.5cm}
	\end{minipage}%
	\hfill
	\begin{minipage}[t]{0.49\linewidth}
		\centering
		\includegraphics[width=3.2in]{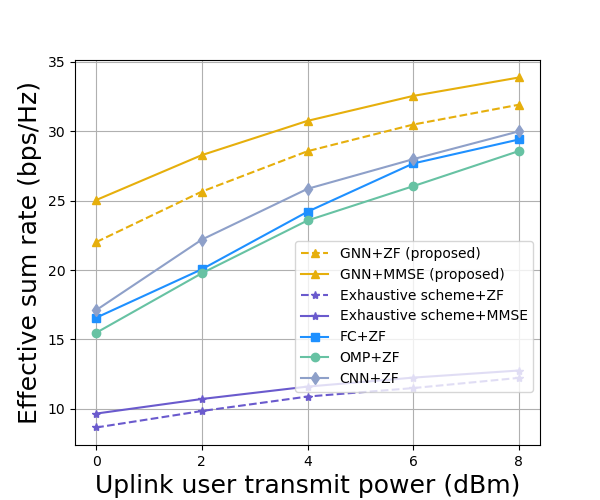}
		\vspace{-0.2cm}
		\caption{Effective sum rate of different schemes as function of the uplink user transmit power. }
		\label{up_eff_sumrate}\vspace{-0.5cm}
	\end{minipage}%
	\hfill
\end{figure*}

In Fig. \ref{up_sumrate}, we compare the sum rate performance of the proposed scheme and the benchmark schemes for different UL user transmit powers. Note that the curves with ``+ZF" and ``+MMSE" in the label indicate that the ZF and the MMSE criteria were used for the design of the digital TPC, respectively. Firstly, Fig. \ref{up_sumrate} shows that our proposed GNN-based multi-user DL transmission scheme can approach the performance of the near-field exhaustive search for all considered UL transmit powers, yet our proposed scheme causes an extremely low pilot overhead.  Secondly, our proposed scheme exhibits significantly better performance than the other benchmark schemes in terms of the sum rate. Although all schemes are based on neural networks, the GNN in the proposed scheme can exploit the pilot signals of the surrounding users by virtue of its unique ``aggregation" and ``combination" structure, thus outperforming both the FC network-based and CNN-based schemes in terms of sum rate. The OMP-based scheme struggles to achieve high sum rate performance, because the interference between the users cannot be fully eliminated by the hybrid TPC for the OMP-based scheme. In addition, we also compare the performance of the proposed scheme for ZF and MMSE digital TPCs. As expected, the performance of the proposed scheme with the MMSE TPC is better than that with the ZF TPC. 

\begin{figure*}
	\begin{minipage}[t]{0.49\linewidth}
		\centering
		\includegraphics[width=3.2in]{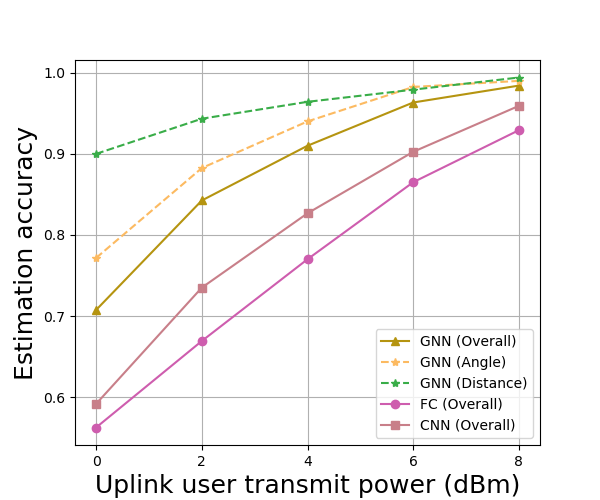}
		\vspace{-0.2cm}
		\caption{Estimation accuracy of different schemes. }
		\label{up_acc}
	\end{minipage}%
	\hfill
	\begin{minipage}[t]{0.49\linewidth}
		\centering
	\includegraphics[width=3.2in]{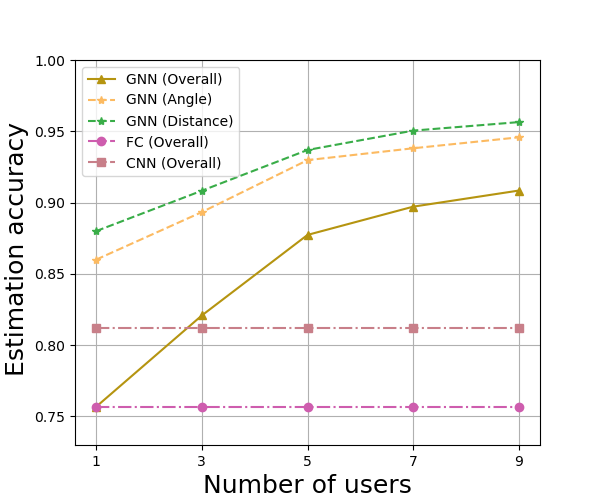}
	\vspace{-0.2cm}
	\caption{Estimation accuracy of  different numbers of users.  }
	\label{usernum_acc}\vspace{-0.5cm}
	\end{minipage}%
	\hfill
	\vspace{-0.3cm}
\end{figure*}

In Fig. \ref{up_eff_sumrate}, we compare the performance of the considered schemes in terms of the effective sum rate for different uplink transmit powers. Comparing Figs. \ref{up_sumrate} and \ref{up_eff_sumrate}, we can see that although the exhaustive near-field search scheme performs better in terms of sum rate, it is quite unsatisfactory in terms of the effective sum rate. This is due to the fact that the exhaustive search introduces an excessive pilot overhead for near-field channels. Specifically, for the given simulation conditions, the exhaustive near-field scheme has to transmit $N_{\mathrm{BS} }S=1280$ pilot symbols in the UL, which reduces the time for DL data transmission in a communication period and ultimately leads to a drop in the effective sum rate. By contrast, our proposed scheme only has to transmit $N_{\mathrm{BS} }/M+2K=80$ pilot symbols, which reduces the pilot overhead by more than 93\% compared to the exhaustive near-field search scheme. Therefore, it can be concluded that our proposed scheme not only achieves a high sum rate but also causes a low pilot overhead. As for the benchmark schemes other than exhaustive search, we set their pilot overhead to the same value as that of the proposed scheme. As can be seen, the proposed scheme achieves higher sum rates and higher effective sum rates for the same pilot overhead.

In Fig. \ref{up_acc}, we present the performance of the proposed GNN-based networks, i.e., the angle network and the distance network, in terms of estimation accuracy as functions of the UL transmission power. If $\sigma _{k}^{1}$ is the angle index of the optimal near-field codeword for user $k$, then the angle estimate provided by the angle network for user $k$ is correct. A similar definition holds for the distance estimate. Naturally, the overall estimation accuracy of the proposed GNN-based scheme is equal to the product of the estimation accuracies of the angle network and that of the distance network. As the UL transmit power increases, the estimation accuracies of both the angle network and the distance network also gradually improve and become near perfect at higher UL transmit powers, e.g., when $P_{\mathrm{ul} } $ is higher than 6 dBm. In terms of overall estimation accuracy, the proposed GNN outperforms both the FC network and the CNN, especially when the UL transmit power is low. This is attributed to the fact that the GNN can acquire environmental information to help it differentiate between NLOS paths, noise-induced abnormalities, and true LOS paths.

\begin{figure*}
	\vspace{-0.1cm}
	\begin{minipage}[t]{0.45\linewidth}
			\centering
		\includegraphics[width=3.5 in]{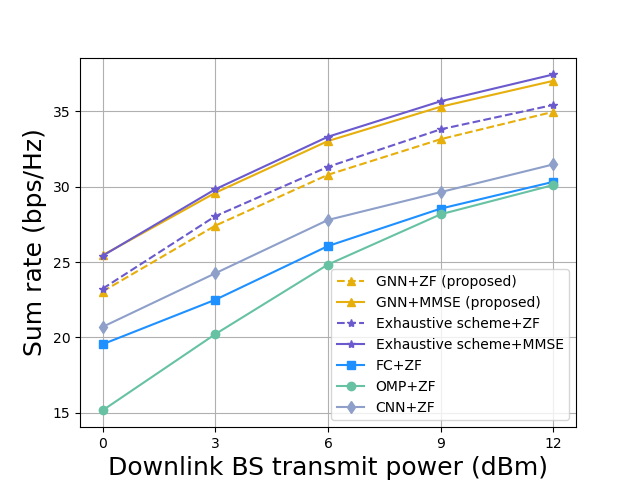}
		\vspace{-0.4cm}
		\caption{Sum rate of different schemes as function of the downlink transmit power. }
		\label{dp_sumrate}
	\end{minipage}%
	\hfill
	\begin{minipage}[t]{0.45\linewidth}
		\centering
		\includegraphics[width=3.2in]{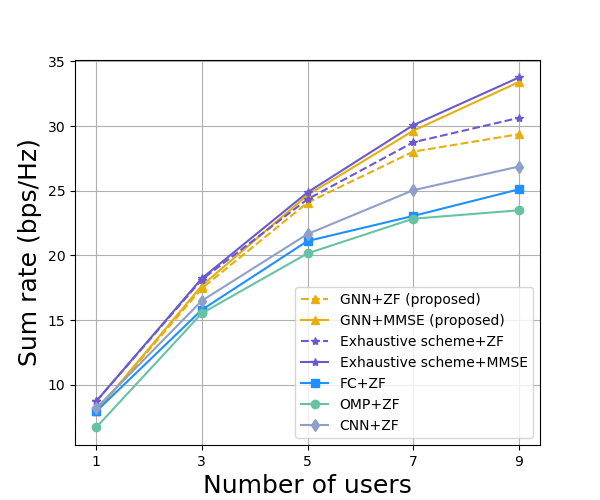}
		\vspace{-0.2cm}
		\caption{Sum rate of considered schemes for different numbers of users.}
		\label{usernum_sumrate}\vspace{-0.5cm}
	\end{minipage}%
	\hfill
	\vspace{-0.2cm}
\end{figure*}

In Fig. \ref{usernum_acc}, we evaluate the performance of the proposed GNN-based scheme for different numbers of users in terms of estimation accuracy. As the number of users increases, the estimation accuracies of both the angle network and of the distance network improve significantly. Actually, as the number of users increases, the GNN can acquire information about the beamforming gains of more surrounding users before estimating the optimal beam for a particular user. This allows the GNN to sense the surrounding environment more accurately and thus to better differentiate between NLOS paths, noise-induced abnormalities, and true LOS paths. However, the estimation accuracies of the FC network and the CNN do not improve, as the number of users increases. This is because, in the FC-network-based and CNN-based schemes, the beam training of different users is carried out in parallel and disjointly, which makes the estimation accuracy of these schemes independent of the number of users. In other words, the structures of the FC network and the CNN prevent them from exploiting the pilot signals of the surrounding users for accurate beam estimation. Furthermore, we also observe from Fig. \ref{usernum_acc} that when the number of users is only 1, i.e., when the multi-user scenario becomes a single-user scenario, the overall estimation accuracy of the GNN degrades to the same level as that of the FC network. The reason for this is that the GNN and the FC network have similar structures and become identical, when the GNN is unable to glean any environmental information through ``aggregation" and ``combination". Overall, our results demonstrate that the GNN is indeed capable of learning the surrounding environment in multi-user scenarios and ultimately achieving a significant performance gain. Moreover, this performance gain becomes more evident as the number of users in the system increases.

In Fig. \ref{dp_sumrate}, we compare the performance of all considered schemes for different DL transmit powers in terms of the sum rate. Similar to the case illustrated in Fig. \ref{up_sumrate}, it can be seen in Fig. \ref{dp_sumrate} that the proposed scheme approaches the exhaustive near-field search scheme in terms of sum rate and simultaneously outperforms the remaining three benchmark schemes for all considered DL transmit powers. Figs. \ref{up_sumrate} and \ref{dp_sumrate} reveal that the proposed scheme is quite robust and can provide substantial performance advantages for different UL and DL transmit powers.

In Fig. \ref{usernum_sumrate}, we compare the performance of the considered schemes for different numbers of users in terms of the sum rate. The proposed scheme can still approach the performance of the exhaustive search scheme and outperform the other benchmark schemes, regardless of the number of users. When the number of users is low, the proposed scheme has no clear advantage. However, as the number of users increases, the GNN in the proposed scheme achieves a huge sum rate improvement by exploiting information about the surrounding environment. Additionally, as a benefit of the beam allocation scheme in Algorithm \ref{alg2}, the proposed scheme and the baseline schemes other than the OMP-based scheme avoid beam conflicts and thus reduce interference between users. Consequently, when the number of users is high, the sum rates of the proposed scheme and the baseline schemes other than the OMP-based scheme still increase significantly with the number of users. By contrast, the OMP-based scheme no longer achieves a significant sum rate increase, when the number of users exceeds 7 due to the severe inter-user interference that occurs at this point.

\vspace{-0.4cm}
\section{Conclusions}\label{conclusion}
A GNN-based beam training scheme was proposed for multi-user XL-MIMO systems to reduce the pilot overhead required for beam training and to improve the spectral efficiency. In the proposed scheme, only far-field wide beams have to be tested for each user and the corresponding beamforming gain information is mapped to the optimal near-field codeword via a GNN, which can exploit the beamforming gain information of all other users for further improving the beam training performance. In addition, a hybrid TPC design based on GNN was proposed to reduce inter-user interference. Our simulation results showed that the proposed scheme outperforms several benchmark schemes and approaches the exhaustive search, even though the pilot overhead is reduced to 7\% compared to the exhaustive search.


\bibliographystyle{IEEEtran}
\vspace{-0.4cm}
\bibliography{myre}


\end{document}